\documentclass[aps,prb,twocolumn,altaffilletter,nolongbibliography,numerical,flushbottom,secnumarabic,superscriptaddress,floatfix,10pt]{revtex4-2}

\usepackage{graphicx}
\usepackage{braket}
\usepackage{booktabs}  
\usepackage{comment}
\usepackage[normalem]{ulem}
\usepackage[title]{appendix}
\usepackage{amsmath,mathtools,amsthm,amssymb,pifont}
\usepackage{physics}
\usepackage[utf8]{inputenc}
\usepackage[american]{babel}
\usepackage{graphicx,xcolor,bbold,titlesec}
\usepackage{MnSymbol}

\usepackage{float}

\usepackage{mathtools}
\usepackage{tcolorbox}

\tcbset{minimalist/.style={
  boxrule=0.4pt,
  colback=white,
  colframe=black,
  sharp corners,
  boxsep=4pt,
  left=4pt,
  right=4pt,
  top=4pt,
  bottom=4pt,
}}

\usepackage{tikz}
\usetikzlibrary{decorations.pathreplacing, calc}

\usepackage[breaklinks, pdftex, hyperfootnotes=true, pdfpagelabels, bookmarks, pageanchor]{hyperref}
\hypersetup{%
	colorlinks=true, linktocpage=true, pdfstartpage=1, pdfstartview=FitH, pdfborder={0 0 0},%
	breaklinks=true, pdfpagemode=UseNone, pageanchor=true, pdfpagemode=UseOutlines,%
	plainpages=false, bookmarksnumbered, bookmarksopen=true, bookmarksopenlevel=1,%
	hypertexnames=true, pdfhighlight=/O,
	urlcolor=red, linkcolor=red, citecolor=red,
	}
\usepackage{orcidlink}
\usepackage{tikz,ifthen}
\usepackage{bbold}
\usepackage{tikz-network}
\usetikzlibrary{patterns,decorations.pathreplacing,calligraphy}
\usetikzlibrary{shapes,arrows,decorations.pathmorphing}

\graphicspath{{pictures}}

\begin{document}

\newcommand{\titleinfo}{ Simulating generalised fluids  via interacting wave packets evolution }
\title{\titleinfo}

\author{Andrew Urilyon~\orcidlink{0000-0001-8960-5388}}
\affiliation{Laboratoire de Physique Th\'eorique et Mod\'elisation, CNRS UMR 8089,
CY Cergy Paris Universit\'e, 95302 Cergy-Pontoise Cedex, France}

\author{Leonardo Biagetti~\orcidlink{0009-0009-4170-8447}}
\affiliation{Laboratoire de Physique Th\'eorique et Mod\'elisation, CNRS UMR 8089,
CY Cergy Paris Universit\'e, 95302 Cergy-Pontoise Cedex, France}

\author{Jitendra Kethepalli~\orcidlink{0000-0001-7326-0231} }
\affiliation{Laboratoire de Physique Th\'eorique et Mod\'elisation, CNRS UMR 8089,
CY Cergy Paris Universit\'e, 95302 Cergy-Pontoise Cedex, France}

\author{Jacopo De Nardis~\orcidlink{0000-0001-7877-0329}}
\affiliation{Laboratoire de Physique Th\'eorique et Mod\'elisation, CNRS UMR 8089,
CY Cergy Paris Universit\'e, 95302 Cergy-Pontoise Cedex, France}

\begin{abstract}
One–dimensional integrable and \emph{quasi-}integrable systems display, on macroscopic scales, a universal form of transport known as \textit{Generalized Hydrodynamics} (GHD).  
In its standard Euler-scale formulation, GHD mirrors the equations of a two-dimensional compressible fluid but ignores fluctuations and becomes numerically unwieldy as soon as integrability-breaking perturbations are introduced. We show that GHD can be efficiently simulated as a gas of \emph{semiclassical wave packets}—a natural generalisation of hard-rod particles—whose trajectories are efficiently mapped onto those of point particles.  
This representation (i) provides a transparent route to incorporate integrability-breaking terms, and (ii) automatically embeds the exact fluctuating-hydrodynamics extension of GHD. The resulting framework enables fast, large-scale simulations of quasi-integrable systems even in the presence of complicated integrability-breaking perturbations.  
It also manifest the pivotal role of two-point correlations in systems confined by external potentials: we demonstrate that situations where local one-point observables appear thermalised can nevertheless sustain long-lived, far-from-equilibrium \emph{long-range} correlations for arbitrarily long times, signaling that, differently from what previously stated, true thermalisation is\textit{ not} reached at diffusive time-scales.  
\end{abstract}

\maketitle

\section{Introduction} 

Generalized hydrodynamics (GHD) \cite{PhysRevX.6.041065, PhysRevLett.117.207201} has emerged as a powerful framework for studying large-scale transport and relaxation in one-dimensional (1D) systems that are either integrable or near-integrable  \cite{PhysRevLett.122.090601,PhysRevLett.119.195301, PhysRevLett.121.160603, de_nardis_diffusion_2019, doyon_lecture_2019, doyon2021free, SciPostPhys.3.6.039, doyon2022diffusion, PhysRevLett.127.130601, Doyon2018, PhysRevLett.122.090601, Bastianello_2022, 10.21468/SciPostPhysCore.3.2.016, bastianello2018sinh, RevModPhys.93.025003, PhysRevLett.122.240606, Bulchandani_2021, Hubner2023, PhysRevResearch.6.013328, Bulchandani_2021,doi:10.1126/science.abf0147,10.21468/SciPostPhysCore.7.2.025,PhysRevLett.124.140603,Besse2023DissipativeGHD, Bulchandani2018BetheBoltzmann, Moller2022Bridging,PhysRevLett.123.130602, Moller2024Anomalous,10.21468/SciPostPhys.9.4.044,Doyon2025Perspective,Bonnemain2022,PhysRevLett.133.107102}, with extensive experimental applications to quasi-one-dimensional cold atomic and condensed matter setups \cite{PhysRevLett.133.113402,PhysRevLett.126.090602,Kao2021,Le2023,PhysRevX.12.041032,PhysRevLett.122.090601,doi:10.1126/science.abf0147,2406.17569,Wei2022,2505.10550,2505.05839}.  While GHD accurately captures the evolution of average densities and currents, it does not fully describe fluctuations and can become numerically cumbersome when including small integrability-breaking perturbations.  Such perturbations, ubiquitous in realistic cold-atom or condensed-matter setups, demand complex analytical and computational corrections to the continuum description.

\begin{figure}[t!]
  \includegraphics[width=0.5\textwidth]{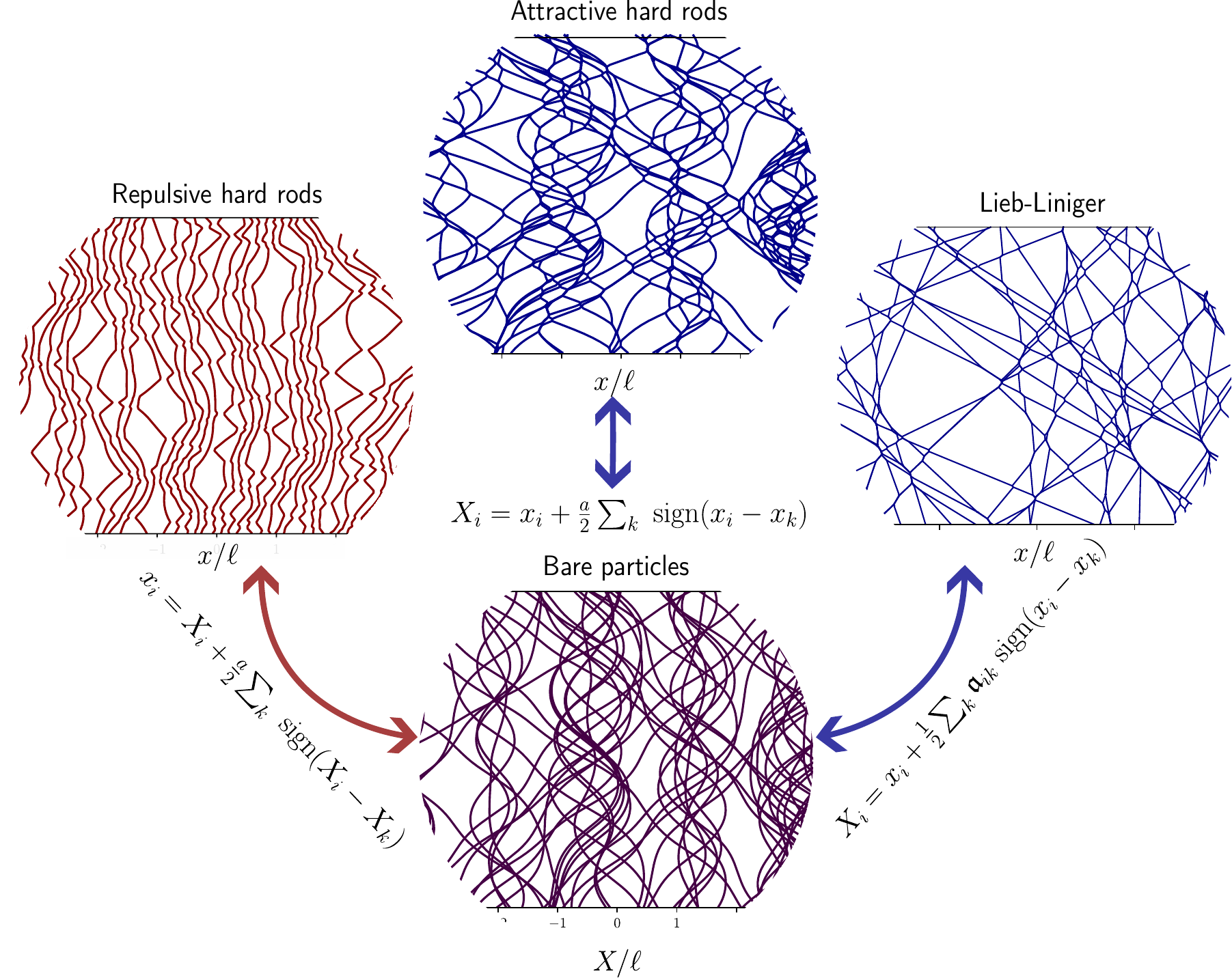} 

  \caption{ Example trajectories of the WPG gas, with coordinates $x$, as obtained by mapping from the bare coordinates $X$. By choosing an appropriate scattering shift $\mathfrak{a}$ the interacting coordinates are realized for different models: pictured are the repulsive hard rod gas $\mathfrak{a}(\theta,\theta') = -1$ (upper left), attractive hard rod gas with $\mathfrak{a}(\theta,\theta') = 1$ (center), and Lieb-Liniger WPG with $\mathfrak{a}(\theta,\theta') = 2c/((\theta-\theta')^2 + c^2)$ and $c=1$ (upper right), all quenched from a homogeneous initial state to a cosine potential with the interacting dynamics determined from the bare coordinates (bottom). Notice how the attractive hard rod gas has similar scattering dynamics as the Lieb-Liniger gas and therefore can be used as a simpler theoretical toy model.  }\label{fig:cartoon}
\end{figure}

In this paper, building on Refs.~\cite{10.21468/SciPostPhys.13.3.072,Bonnemain2025,2503.08018}, and in particular on \cite{Hubner2023,PhysRevLett.132.251602}, we adopt the alternative picture in which the generalized fluid is viewed as a gas of semiclassical point-like wave packets whose trajectories map onto those of noninteracting (``bare'') particles; see Fig.~\ref{fig:cartoon}. This particle-based representation recovers the standard hydrodynamic equations under averaging over initial realisations, but it also goes much beyond by incorporating correlations and offering a transparent way to describe integrability-breaking effects.  Unlike conventional GHD approaches, which become intractable when adding noise and higher-order interactions, the soliton-gas formalism naturally encodes fluctuating hydrodynamics, since interactions appear as kinetic corrections in the bare–interacting mapping.  Consequently, both average properties and fluctuations can be treated on an equal footing.

The bare–interacting soliton-gas mapping thus provides a versatile framework for exploring a broad class of phenomena in 1D fluids.  It smoothly interpolates between strictly integrable models and quasi-integrable regimes dominated by realistic perturbations.  We demonstrate that our approach quantitatively reproduces known GHD results in integrable limits and extends readily to systems with broken integrability (overcoming the issues coming from solving complicated partial differential equations \cite{Mller2023}). Moreover,  we show that when integrable systems are subjected to external trapping potentials, the rapidity (the quasiparticle momentum) distribution appears thermalized on diffusive time-scales (as demonstrated in previous works \cite{PhysRevLett.125.240604,PhysRevResearch.6.023083}); yet, we also show the surprising and remarkable fact that long-range correlations generated during intermediate times remain finite even at very large times.  We argue that intermediate-time ballistic transport builds up these long-range correlations, as first noted in \cite{shortLetterBMFT, bmftdoyon, PhysRevE.111.024141,Biagetti2025,2503.07794}, (and which are also the source of different anomalous fluctuations in integrable systems, see, for example, \cite{GopalakrishnanAnomalous, Gopalakrishnan2024, PhysRevLett.128.090604, PhysRevE.111.015410, PhysRevE.111.024141}) and that integrability-breaking terms need additional time-scales in order to fully dissipate them.  In cases such as harmonic or quartic traps, these correlations can be strong enough to prevent true thermalization also in the momentum distribution.  Our findings thus prompt a reassessment of thermalization in the presence of integrability-breaking. Nevertheless, the relevance of our results is beyond integrable systems, since thermalization in 1D systems is a problem shared by different fields such as plasma physics and astrophysics \cite{Eldridge1963, Dawson1964, Rouet1991, Fouvry2019, Fouvry2020, Rybicki1971, Reidl1988, Milanovi1998, Tsuchiya1996, Miller1996}. 
\\

This paper is organized as follows:
Sec.~\ref{sec: theory_integrable_GHD} provides an introduction on Generalized Hydrodynamics for purely integrable systems; in Sec.~\ref{sec: theory_quasi_integrable_GHD} we show GHD equations in the presence of spatially
inhomogeneous perturbations; in Sec.~\ref{sec: wave_packet_gas} we introduce the wave packet gas formulation of GHD and present the algorithm to implement it in classical and quantum integrability breaking protocols;
in Sec.~\ref{sec: numerics_ext_potentials} we show the numerical analysis for the evolution of WPG in inhomogeneous external traps;
Sec.~\ref{sec: numerics_two_body_pot} explains how WPG algorithm can be implemented for more generic two-body integrability breaking potentials; Sec.~\ref{sec: conclusions} contains our conclusions with an outlook on potential applications of WPG.

\section{ Generalised Hydrodynamics and effective viscosities }
\label{sec: theory_integrable_GHD}
GHD is known to be the correct hydrodynamic theory in the context of 1D quantum integrable systems, where it describes the non-equilibrium evolution of density fields of conserved charges, provided that the inhomogeneity is sufficiently smooth.
There is a particularly natural way to understand GHD by beginning with the continuity equation for the aforementioned local density fields and associated local current 
\begin{eqnarray}
    \partial_t q_i(x,t) + \partial_x j_i(x,t) = 0 .
\end{eqnarray}
These local density fields are defined within fluid cells, assumed to have a hydrodynamic length scale $\ell$, that have relaxed such that around their space-time positions, scaled by this hydrodynamic length, these cells are relaxed. Note that $\ell$ sets the length scale, however, by appropriate rescaling of $x$, $t$ this scale will be exactly canceled and the hydrodynamic equation of the corresponding scale follows.
Within each of these fluid cells, a probability distribution is approximately given by a slowly varying generalized Gibbs ensemble (GGE)
\begin{eqnarray}\label{eq:hydrostate}
    \rho \sim \exp\!\left[ - \ell \int \! dx \, \beta^{i}(x,t)\,q_i(x) \right] ,
\end{eqnarray}
where from now on we shall use the repeated-index convention.  
From this \textit{fluid-cell assumption}, a \textit{hydrodynamic expansion} can be performed, which, provided the generalized temperature is sufficiently smooth, determines the current as a functional of the averages of all the conserved charge densities $\langle q_k(x,t) \rangle$,
\begin{eqnarray}
    \langle j_i(x,t) \rangle = j_i[\vec q](x,t) + \frac{\delta j_i}{\delta \beta^j}\bigg|_{[\vec q]} \partial_x \langle q_j(x,t) \rangle + \dots\,.
\end{eqnarray}
At zeroth order, $\mathcal{O}(\ell^0)$, this yields Euler GHD, and at $\mathcal{O}(\ell^{-1})$ a Navier–Stokes-type correction associated with diffusive GHD can be found~\cite{PhysRevLett.121.160603,PhysRevB.98.220303}.  
Importantly, Euler GHD in the presence of external trapping potentials lacks any mechanism for thermalisation (which is expected as the underlying integrability of the system is broken) and requires the presence of the Navier–Stokes terms before any thermalisation can be observed~\cite{PhysRevLett.125.240604,Durnin2021,PhysRevResearch.6.023083}. Namely, it is the combination of internal viscosities and external breaking of integrability (via external forces, for example) that induces thermalization.

Although diffusive GHD successfully captures certain integrability-breaking relaxation dynamics, the Navier–Stokes derivation from the hydrodynamic expansion relies on the assumption that no long-range correlations can develop between fluid cells.  
This is valid at very short times, within the linear-response regime; however, at intermediate times long-range correlations are known to develop~\cite{shortLetterBMFT,bmftdoyon}.  
These transient long-range correlations can be derived using the cumulant expansion, as described in Ref.~\cite{hubner2024diffusive,2503.07794}. There, it was shown that typically integrable systems fail to locally equilibrate, invalidating the applicability of the hydrodynamic states~\eqref{eq:hydrostate}.  
Therefore the correct equilibrium states are those that retain non-trivial information about two-point functions,
\begin{multline}
  \rho_{\text{LR}}
  = Z^{-1}\,
    \exp\!\Bigl[
      - \ell \!\int\! dx\,\beta_{(1)}^{i}(x)\,q_i(x)
      \\ - \ell \!\int\! dx\,dx'\,
        \beta_{(2)}^{ij}(x,x')\,E_{ij}(x,x')
    \Bigr],
  \label{eq:LRstate}
\end{multline}
where $E_{ij}(x,x')\sim\ell^{-1}$ are the regularised (namely, with the divergent part in $x=x'$ subtracted out) two-point correlations of the hydrodynamic charges; namely, the dynamical correlation functions are composed of a local $\delta$-peaked component, whose amplitide is given by the local susceptibility matrix $C_{ij}$, and a regular long-range part which we denote as $E_{ij}$, namely
\begin{eqnarray}
    S_{ij}(x,y,t) = \delta(x-y) C_{ij}(x,t) + E_{ij}(x,y,t) .
\end{eqnarray}
Using the local-equilibrium states~\eqref{eq:LRstate} one can expand the current to include $1/\ell$ corrections and derive the hydrodynamic equation
\begin{eqnarray}
\label{eq:diffusion_full1}
    \partial_t q_i + A^k_i \partial_x q_k &=& -\partial_x \left( \frac{1}{2\ell} \frac{\delta^2 j_i}{\delta q_r \delta q_k}\bigg|_{\vec q} E^{\text{sym}}_{rk}(x) \right) \nonumber \\
    &\overset{t\ll 1}{\approx}& \frac{1}{2\ell}\,\partial_x \!\left( \mathcal{D}_i^{\;k}\,\partial_x q_k \right) ,
\end{eqnarray}
with the point-splitting defined as $E^{\text{sym}}_{ij}(x)\equiv (E_{ij}(x^+,x^-)+E_{ij}(x^-,x^+))/2$.  
At intermediate times the regular correlation must be evolved simultaneously, according to
\begin{eqnarray}
    \label{eq:diffusion_full2}
    \partial_t E_{ij} + \partial_x (A_i^{\;k} E_{kj}) + \partial_y (E_{ik} A_j^{\;k})
    \nonumber\\
    = - \delta(x-y)\,\langle j^-_h, q_i, q_j \rangle C^{hk}\partial_x q_k .
\end{eqnarray}
Evolution at the diffusive scale from an initial locally equilibrated GGE state thus requires the long-range correlation functions to be evolved simultaneously, making the dynamics substantially more complicated than those of Euler GHD.

It is often convenient to recast the hydrodynamic description in terms of quasiparticles instead of local charges.  
A charge density $q_i(x)$ can be written in terms of the quasiparticle density $\rho(\theta,x)$ as
\begin{equation}
  q_i(x) = \int d\theta\,h_i(\theta)\,\rho(\theta,x),
  \label{eq:charge2qp}
\end{equation}
where the form factors $\{h_i(\theta)\}$ form a complete basis, so that Eq.~\eqref{eq:charge2qp} is invertible.  
Quasiparticles may be regarded as traveling wave packets (see the following sections) whose group velocity $v^{\text{bare}}_\theta$ is a continuous function of the rapidity~$\theta$.  
Their dynamics involve only elastic two-body scattering events, characterised by the phase shift $\varphi_{\theta,\theta'}$.  The bare velocity $v^{\text{bare}}_\theta$, together with the bare momentum $k_\theta$ and the scattering phase, fully specifies the microscopic model at hand.

In the quasiparticle representation, the naive Navier–Stokes equation reads
\begin{equation}
    \partial_t \rho_\theta + \partial_x (v^{\text{eff}}_\theta \rho_\theta)
    = \frac{1}{2}\,\partial_x \bigl[ \mathfrak{D}\,\partial_x \rho \bigr]_\theta ,
\end{equation}
where the effective velocity and the diffusion constant are given in terms of the local quasiparticle density,
\begin{eqnarray}
    v^{\text{eff}}_\theta &=& v^{\text{bare}}_\theta + \int d\mu\,\mathfrak{a}_{\theta\mu}\,\rho_\mu\,(v^{\text{eff}}_\mu - v^{\text{eff}}_\theta) , \\
    \mathfrak{D}_{\alpha\gamma}
    &=& R^{-t}_{\alpha\theta}\,
        \frac{\delta_{\theta\mu}\bigl[\int d\lambda\,W_{\lambda\theta}\bigr] - W_{\theta\mu}}
             {\rho_{T,\theta}\rho_{T,\mu}}\,
        R^{t}_{\mu\gamma} ,                         \\
    W_{\theta\mu} &=& \rho_\theta\,f_\theta\,\bigl[\varphi^{\text{dr}}_{\theta\mu}\bigr]^2
                     \bigl|v^{\text{eff}}_\theta - v^{\text{eff}}_\mu\bigr| ,
\end{eqnarray}
with $R_{\alpha \theta}\equiv \delta_{\alpha \theta}-\varphi_{\alpha \theta} n_\theta$ and the rescaled scattering shift $\mathfrak{a}_{\theta\mu}=2\pi\varphi_{\theta\mu}/k'_\theta$.  
Here $\rho_{T,\theta}=k'_\theta/(2\pi)+\int d\mu\,\varphi_{\theta\mu}\rho_\mu$ is the total density, $g^{\text{dr}}_\theta=g_\theta+\int d\mu\,\varphi_{\theta\mu}n_\mu g^{\text{dr}}_\mu$ denotes dressing, $n_\theta=\rho_\theta/\rho_{T,\theta}$ is the occupation function, and $f_\theta=\{1,\,1-n_\theta,1+n_\theta\}$ is the statistical factor (classical, Fermi-Dirac or Bose-Einstein, respectively).  
As specified above, the Navier–Stokes equation is only an approximation, valid at short times, whereas the correct diffusive dynamics are described by the two coupled equations~\eqref{eq:diffusion_full1} and~\eqref{eq:diffusion_full2}, which, in quasiparticles, read
\begin{eqnarray}
\label{eq:diffusion_full1_qp}
    \partial_t \rho_\theta
    + \partial_x (v^{\text{eff}}_\theta \rho_\theta)
    &=& \frac{1}{2}\,\partial_x \!\bigl( H_\theta^{\alpha\beta}\,E^{\text{sym}}_{\alpha\beta}(x) \bigr) ,
    \\[4pt]
\label{eq:diffusion_full2_qp}
    \partial_t E_{\theta\theta'}(x,x')
    &+& \partial_x (A_\theta^{\;\alpha} E_{\alpha\theta'})
        + \partial_{x'} (A_{\theta'}^{\;\alpha} E_{\theta\alpha})
    \nonumber\\
    &=& \delta(x-x')\,M_{\theta\theta'}^{\;\alpha}\,\partial_x \rho_\alpha ,
\end{eqnarray}
with repeated indices integrated.  
The tensor operators are
\begin{eqnarray}
    A_\theta^{\;\alpha}
      &=& R^{-t}_{\theta\mu}\,v^{\text{eff}}_\mu\,R^{t}_{\mu\alpha} ,         \\
    M_{\theta\theta'}^{\;\alpha}
      &=& \bigl[ R^{-t}_{\theta\mu} R^{-t}_{\theta'\gamma}\,
                 \varphi^{\text{dr}}_{\mu\gamma}\,
                 \tfrac{\rho_\gamma f_\gamma}{\rho_{T,\mu}}\,
                 (v^{\text{eff}}_\mu - v^{\text{eff}}_\gamma)\,
                 R^{t}_{\mu\alpha} \bigr]_{(\theta,\theta')} ,               \\
    H_\theta^{\alpha\beta}
      &=& \bigl[ R^{-t}_{\theta\gamma}\,
                 (v^{\text{eff}}_\mu - v^{\text{eff}}_\gamma)\,
                 \tfrac{\varphi^{\text{dr}}_{\gamma\mu}}{\rho_{T,\gamma}}\,
                 R^{t}_{\mu\alpha} R^{t}_{\gamma\beta} \bigr]_{(\alpha,\beta)} ,
\end{eqnarray}
where $[\bullet]_{(\theta,\theta')}$ denotes the sum over permutations of the indices $\theta,\theta'$.  
The inclusion of two-point correlations is crucial for accurately capturing hydrodynamic evolution up to diffusive time scales.  
As demonstrated in Ref.~\cite{hubner2024diffusive,2503.07794}, \textit{the Navier–Stokes equation applies only in the linear-response regime}—i.e.\ for hydrodynamic evolution of initial states weakly perturbed from equilibrium—\textit{or at very short times}, when the long-range part $E_{\theta\theta'}(x,x')$ can be neglected.  
However, \textit{in generic systems that support ballistic modes—provided the mode velocity depends non-linearly on the local state—hydrodynamic evolution generates long-range correlations that do not decay in time}. 
As argued below, the same remains true in the presence of weak integrability breaking, such as that induced by a trapping potential, making the inclusion of long-range correlations necessary in those cases as well.

\section{Trap-induced integrability breaking and thermalisation}
\label{sec: theory_quasi_integrable_GHD}
As shown in Refs.~\cite{PhysRevLett.125.240604,doyon_note_2017,Durnin2021,PhysRevResearch.6.023083}, starting from a generic integrable Hamiltonian $H_{0}$ one may introduce a spatially inhomogeneous perturbation of the form
\begin{equation}
  H = H_{0} + \int dx\,V(x)\,q_{i_{0}}(x),
  \label{eq:H_inhom}
\end{equation}
where $q_{i_{0}}(x)$ is an arbitrary conserved charge density and $V(x)$ is a smooth external potential.

Under this perturbation quasiparticles not only drift in real space but also acquire a flow in rapidity (momentum) space.  
Their current can be written as a two-dimensional vector in this space \cite{doyon_note_2017},
\begin{equation}
  \mathbf{J} = \bigl( v^{\text{eff}}_\theta \rho_\theta ,\,
                       a^{\text{eff}}_\theta \rho_\theta \bigr) ,
  \label{eq:J_theta}
\end{equation}
where $a^{\rm eff}$ is defined analogously as the effective velocity, bare velocity replaced by the
bare acceleration
$a^{\text{bare}} = -h_{i_{0}}\,\partial_x V(x)$.
Because transport now proceeds in both real and rapidity space, the Euler–Navier–Stokes equation generalizes to
\begin{equation}
  \partial_t \rho
  + \boldsymbol{\nabla}\!\cdot\!\mathbf{J}
  = \frac{1}{2}\,
    \boldsymbol{\nabla}\!\cdot\!
    \bigl( \boldsymbol{\mathfrak{D}}\,
           \boldsymbol{\nabla}\rho \bigr),
  \qquad
  \boldsymbol{\nabla} \equiv (\partial_x,\partial_\theta),
  \label{eq:nabla_eq}
\end{equation}
where $\boldsymbol{\mathfrak{D}}$ denotes the rapidity-resolved diffusion tensor.

Equation~\eqref{eq:nabla_eq} possesses the thermal local-density-approximation Gibbs ensemble as a fixed point of evolution, namely the thermal state modulated at each point by the external potential,
\begin{equation}
  \rho_{\text{th}}(x) = Z^{-1}\,\exp\!\bigl[ -\beta\,( H_{0} + V(x)\,q_{i_{0}}(x) - \mu N ) \bigr] .
\end{equation}
Although a formal proof is still missing, it is widely believed to be the unique steady state, apart from exceptional choices of $V(x)$.  
Accordingly, integrable systems subjected to external potentials are expected to thermalise on diffusive time scales—a prediction borne out numerically except for a few notable deviations in trapped hard-rod gases \cite{PhysRevLett.120.164101,PhysRevE.108.064130}.

However, recent developments have shown that the Navier–Stokes equation~\eqref{eq:nabla_eq} is not correct on pre-thermalisation time scales \cite{PhysRevResearch.6.023083}.
We therefore have to formulate the problem including two-point correlation functions.  
Below we derive coupled evolution equations for the density and its connected two-point function.
They coincide with those in the previous section, but include the acceleration terms induced by the external force field
\begin{eqnarray}
\label{eq:diffusion_full1_qp_trap}
    \partial_t \rho_\theta
    &&+ \partial_x (v^{\text{eff}}_\theta \rho_\theta)
    + \partial_\theta (a^{\text{eff}}_\theta \rho_\theta)=
    \nonumber\\
    &&= \frac{1}{2}\,\partial_x \!\bigl( H_\theta^{\alpha\beta} E^{\text{sym}}_{\alpha\beta} \bigr)+ \frac{1}{2}\,\partial_\theta \!\bigl( H_\theta^{\mathfrak{f},\alpha\beta} E^{\text{sym}}_{\alpha\beta} \bigr)
\end{eqnarray}
\begin{eqnarray}
\label{eq:diffusion_full2_qp_trap}
    \partial_t E_{\theta\theta'}(x,x')
    &&+ \partial_x (A_\theta^{\;\alpha} E_{\alpha\theta'})
    + \partial_{x'} (A_{\theta'}^{\;\alpha} E_{\theta\alpha})+
    \nonumber\\
    &&+ \partial_\theta (A_\theta^{\mathfrak{f},\gamma} E_{\gamma\theta'})
    + \partial_{\theta'} (A_{\theta'}^{\mathfrak{f},\gamma} E_{\theta\gamma})=
    \nonumber\\
    &&=
    \delta(x-x')\bigl( M_{\theta\theta'}^{\;\alpha} \partial_x \rho_\alpha
    + M_{\theta\theta'}^{\mathfrak{f},\alpha} \partial_\alpha \rho_\alpha \bigr)\,.
\end{eqnarray}
The matrix definitions are
\begin{eqnarray}
\label{eq: A_f matrix}
    A_\theta^{\mathfrak{f},\alpha}
      &=& R^{-t}_{\theta\mu}\,a^{\text{eff}}_\mu\,R^{t}_{\mu\alpha},              \\
\label{eq: M_f matrix}    M_{\theta\theta'}^{\mathfrak{f},\alpha}
      &=& \bigl[ R^{-t}_{\theta\mu} R^{-t}_{\theta'\gamma}\,
    \varphi^{\text{dr}}_{\mu\gamma}\,
    \tfrac{\rho_\gamma f_\gamma}{\rho_{T,\mu}}\,
    (a^{\text{eff}}_\mu - a^{\text{eff}}_\gamma)\,
    R^{t}_{\mu\alpha} \bigr]_{(\theta,\theta')},\\
\label{eq: H_f matrix}    H_\theta^{\mathfrak{f},\alpha\beta}
      &=& \bigl[ R^{-t}_{\theta\gamma}\,
    (a^{\text{eff}}_\mu - a^{\text{eff}}_\gamma)\,
    \tfrac{\varphi^{\text{dr}}_{\gamma\mu}}{\rho_{T,\gamma}}\,
R^{t}_{\mu\alpha} R^{t}_{\gamma\beta} \bigr]_{(\alpha,\beta)} .
\end{eqnarray}
In Appendix \ref{sec: app_diffusive_inhomogeneous} we explicitly derive Eq.~\eqref{eq:diffusion_full1_qp_trap} and~\eqref{eq:diffusion_full2_qp_trap} and show that, for uncorrelated initial states $E(t=0)=0$, in the small time limit $t\to0^+$ they reproduce Eq.~\eqref{eq:nabla_eq}.

Two remarks are necessary.  
First, a faithful simulation of the system requires evolving the fully inhomogeneous correlation function, which is numerically challenging.  
Second, one should ask about the fate of thermalisation in this system of coupled equations.  
We cannot provide a definitive answer here and leave it as an open question.  
However, since there is no term that dissipates correlations, the system cannot thermalise at order $O(\ell^2)$, contrary to what the Navier–Stokes equation would suggest.  
In fact, full thermalisation of the two-point functions appears only at dispersive order in the hydrodynamic expansion, i.e.\ on time scales $\sim O(\ell^3)$.  
More precisely, the relaxation of the two-point function involves three-point functions \cite{Biagetti2025} and can be captured by a diffusive contribution proportional to 
\begin{equation}
\label{eq: diff_correction_correlations}
   \partial_t E_{\theta\theta'} \propto
   H_\theta^{\alpha\beta} E^{\text{sym}}_{\alpha\beta\theta'}(x^+,x^-,x') + \dots ,
\end{equation}
where $E^{\text{sym}}_{\alpha\beta\theta'}$ is the three-point function, symmetrised with respect to the first two spatial variables.  
In this hierarchical structure, the three-point function $E^{\text{sym}}_{\alpha\beta\theta'}$ is expected to evolve through an equation coupled to the first two layers of the hierarchy.  
The resulting cascade of correlations is clearly too complex to numerically simulate, the following section provides a numerical method that circumvents the problem by simulating the effective wave-packet gas, which provides the correct discretisation and contains by definition the correct correlations' dynamics.  

\section{Wave packet gas description of GHD} 
\label{sec: wave_packet_gas}
As we have seen in the previous section, a complete solution of the Generalized hydrodynamics (GHD) including diffusive effects that is capable of describing a system's thermalisation would require solving two coupled equations, one of which describes two-point functions. From a numerical standpoint, the two-point function is quite a complex object. Therefore, especially to go beyond the realm of strictly integrable systems and to be able to efficiently incorporate integrability-breaking perturbations, it is desirable to introduce a particle model whose averages over realizations coincide with those of integrable models in the hydrodynamic limit.

Following recent works, \cite{PhysRevLett.132.251602,10.21468/SciPostPhys.13.3.072,Bonnemain2025,2503.08018,Hubner2023} we introduce the following gas of wave packets (WPG) for this purpose. First, we observe that for a simple model such as the repulsive hard rods, the wave packets coincide with the rods themselves. For a generic interacting integrable model, the WPG is simply described by a gas of interacting solitonic-like particles that scatter via the model-dependent scattering shift. Although another effective particle gas was already introduced in \cite{PhysRevLett.120.045301}, known as the ``flea gas'',  the latter missed the correct hydrodynamic fluctuations. Our  method instead generalizes similar approaches already employed for evolving the repulsive hard rod gas to more general settings and models, and, as we shall show, guarantees the correct viscosities, thermalisation times under integrability breaking, and correlations.



\subsection{The classical hard rods gas}
To elucidate our method, we consider a classical hard rod gas, consisting of either attractive or repulsive particles. In the repulsive case this is the most familiar form of a hard rod gas \cite{PhysRev.171.224,Percus_HR_1969,Boldrighini1983,Spohn1991,Boldrighini1997, PhysRevE.104.064124}, whose extended particles have a finite length $a>0$ that are unable to overlap and instead bounce off of one another exchanging momenta whenever their edges come into contact. In the attractive case the particles all have regions around their center such that all neighbors within a distance $|a|$ away are pulled into the center, as a consequence when a collision occurs these attractive particles will stick together for a period of time based on the relative speed of the colliding particles. The position of the $i$-th particle, $x_i$, is transformed to a corresponding bare particle position, $X_i$, by use of the transformation
\begin{eqnarray}
\label{eq:free_to_int_rWPG}
    X_i = x_i \mp \frac{a }{2} \sum_k {\rm sign}(x_i - x_k) . 
\end{eqnarray}
with the sign corresponding to repulsive ($-$) and attractive ($+$) case, respectively, and where the bare particles coordinates are given by free ballistic propagation
\begin{equation}
    X_i(t) = X_i(0) + \theta_i \ t .
\end{equation}
For the repulsive case this transformation is easily inverted, since the bare particle positions $\{X_i\}$ and WPG $\{ x_i\}$ coordinates are identically ordered, which implies ${\rm sign}(x_i - x_k) = {\rm sign}(X_i - X_k)$. Hence we can write 
\begin{eqnarray}
\label{eq:free_to_int_rWPG_repHR}
    x_i = X_i + \frac{a}{2} \sum_k {\rm sign}(X_i -X_k)\quad({\rm repulsive}\,\,{\rm HR})\,.
\end{eqnarray}
For attractive hard rods, a naive inversion will result in cases with bare coordinates $X_i < X_k$ and interacting coordinates $x_i > x_k$, hence differently ordered, whenever two particles are too close together. Thus, additional care is needed here to keep the mapping consistent. As a resolution, whenever wave packets sit too close to one another they merge into a single position in an example of clustering. This behavior is also observed for more general WPGs, where the trajectories stick together before separating at a later time.

While the dynamics given by \eqref{eq:free_to_int_rWPG} is deterministic, one then needs to sample over an ensemble of initial states in order to describe a thermodynamic state. However the advantage of working with  \eqref{eq:free_to_int_rWPG} is crucial: one does not need to know or use the invariant or thermal distribution of interacting rods, but one can simply initialise the bare particles variables $X_i,\theta_i$, whose statistics is known to be simple Poissoinian, and then perform the mapping to interacting coordinates. We shall show below the algorithm to efficiently sample generic classical hard rods particles.

To make use of the transformation for attractive hard rods the transformation must be inverted, a gradient transform can be implemented; however, for the attractive hard rods a series of inequalities can also be used to determine the interacting positions from the bare coordinates. The key to this method is the introduction of a naive transformation, where ${\rm sign}(x_i - x_j) \to {\rm sign}(X_i - X_j)$, recalled to be the correct inversion for repulsive hard rods, although generally incorrect in the attractive case. In addition to this naive transformation, the ordering of the bare positions $X_j < X_k$ implies $x_j \leq x_k$ when $ x_j = x_k$ these particles are said to be within a cluster. With this in mind it is possible to identify the positions of clusters by finding where this weak ordering fails after a naive transformation. 

Concretely, the bare coordinates $X_{I_1} < \dots < X_{I_m}$ belonging to a single cluster are ordered as $I_1 < \dots < I_m$. Surprisingly, but usefully, the average position of clustered particles in both the true and naive transformations are identical. As a consequence the naive transformation implies the following condition on the first, $x^{\rm naive}_{I_1}$, and last particle positions, $x^{\rm naive}_{I_m}$, within a cluster of $m$-particles relative to the true cluster position $x_I$
\begin{eqnarray}
    x^{\rm naive}_{I_m} < x_I < x^{\rm naive}_{I_1}.
\end{eqnarray}
This condition can be used to group the particles into clusters by hand, which effectively inverts the transformation for the attractive hard rod gas. With this approach numerical solutions could be reproduced more quickly and stably for the attractive hard rod gas.

\subsection{Generic integrable models and quantum statistics}

Generic integrable systems can be viewed as deformations of the hard-rod gas.  
In particular, they are characterised by effective quasi particles whose large-scale generalised hydrodynamics coincide with that of hard rods under the substitution
\begin{equation}
  a \;\longrightarrow\; \mathfrak{a}_{ij} = \frac{2\pi\,\varphi(\theta_i-\theta_j)}{k'(\theta_i)}.
\end{equation}
Some relevant examples, especially its experimental applications, are provided by the Lieb-Liniger model, which describe interacting (quantum) bosonic atoms~\cite{lieb1963,cazalilla_2004,cazalilla_2011,Bouchoule_2022} with Hamiltonian given by
\begin{equation}\label{eq:LLin}
    H_{\rm Lieb-Liniger} =  \sum_j \frac{p_j^2}{2m} +2 c \sum_{i<j}  \delta(x_i-x_j),
\end{equation}
and associated scattering shift $\mathfrak{a}_{ij} = 2c/( (\theta_i -\theta_j)^2 + c^2)$, or the celebrated Toda model \cite{ford1973integrability, toda1967vibration}, for its numerous studies in classical integrability-to-chaos transitions \cite{PhysRevLett.125.040604,PhysRevE.99.022146}, with Hamiltonian
\begin{equation}\label{eq:Toda}
    H_{\rm Toda} =  \sum_j \frac{p_j^2}{2m} +  \sum_{j=1}^{N-1} e^{-(x_{j+1} - x_{j})},
\end{equation}
and associated scattering shift $\mathfrak{a}_{ij} = -2 \log(\theta_i-\theta_j)$. We also set the mass as $m=1$.

The hard-rod mapping of Eq.~\eqref{eq:free_to_int_rWPG} then generalises to generic models, relating interacting and bare particle coordinates through
\begin{equation}
\label{eq:mapping00}
\boxed{  X_i \;=\; x_i
    \;+\; \frac{1}{2}\sum_{k=1}^{N}
     \mathfrak{a}_{ik}\,
      \mathrm{sign}(x_i-x_k), \quad \quad \dot{X}_i = v^{\rm bare}(\theta_i), }
\end{equation}
where the bare velocity parameterization $v^{\rm bare}_\theta$ can account for different types of models (in generic Galilean invariant models one typically has $v^{\rm bare}_\theta=\theta$). This equation was first introduced in \cite{Hubner2023,PhysRevLett.132.251602}. We shall show in Appendix \ref{app:ghdmicro} the argument showing that the dynamics induced by \eqref{eq:mapping00} leads indeed to GHD type of hydrodynamics (notice that a similar argument is also provided in \cite{PhysRevLett.132.251602,2503.08018}).

A crucial difference compared to the hard rods mapping is that it cannot be inverted explicitly. Therefore, in order to invert the transformation a convex surface defined by the phase space position of each particle, $(\vec X, \vec \theta)$, is introduced such that the interacting positions are those positions that minimize the surface
\begin{eqnarray}
        S_{\vec X, \vec \theta}(\vec x) &=& \sum_i \frac{(x_i - X_i)^2}{2} + \frac{1}{2} \sum_{i,j} \mathfrak{a}_{ij} \,  |x_i - x_j| ,\nonumber\\
    \vec x(t) &=& {\rm arg \, min} \, S_{\vec X(t), \vec \theta} (\vec x ).
\end{eqnarray}
Numerically, a gradient descent method is capable of carrying out this minimization, thus identifying a set of interacting coordinates from a set of bare coordinates. By inverting this formula an initial fluctuating state of bare particles is mapped to an initial fluctuating state of a general quantum integrable system. Averages over these initial states reproduce the correct expectation values and correlation functions. 

A less complete particle model of GHD has been realized in the literature prior to both this work and Ref.~\cite{PhysRevLett.132.251602}. In Ref. \cite{PhysRevLett.120.045301} the authors introduced the so-called flea gas algorithm, where hard rod-like particles scatter when they find themselves at distance $\mathfrak{a}_{ij}$. However, such an algorithm does not properly treat multi-scattering processes and, even if it gives the correct average effective  velocities \cite{10.21468/SciPostPhys.8.4.055,PhysRevLett.120.045301,10.21468/SciPostPhys.6.6.070},  it fails to give the correct diffusive terms as both initial fluctuations and correct dynamics are not properly described. 

\begin{figure}
\includegraphics[width=0.45\textwidth]{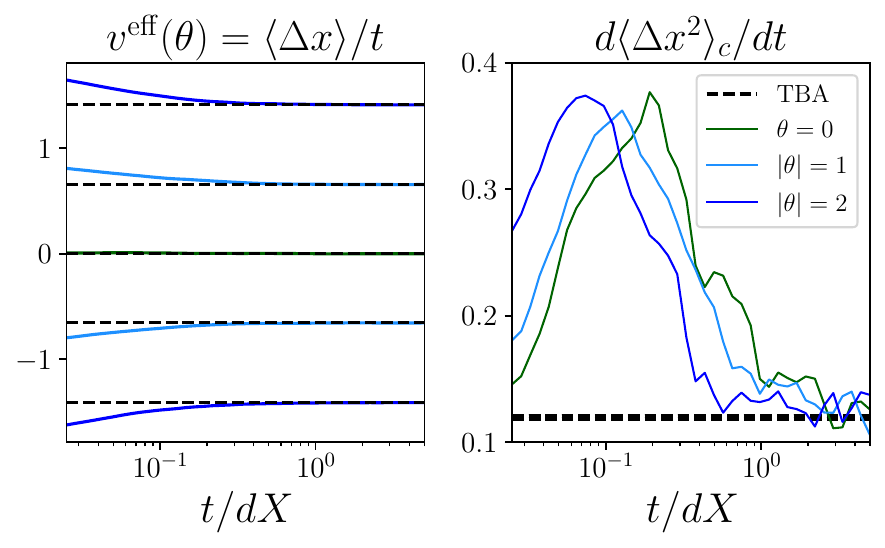}
  \caption{ Evolution of a homogeneous WPG sampled with Fermi-Dirac statistics using the scattering kernel of the Lieb-Liniger model with $\beta = c = \mu = 1$ and with $dX = 40$.  A tracing particle is placed in the center of the system with $\theta = -2,-1,0,1,2$ and the displacement $\Delta x = x(t) - x(0)$ of this particle from its initial position is considered (left) the average displacement $\langle x \rangle$ approaches the corresponding effective velocity from GHD (right) the second moment of displacement approaches the diffusive constant predicted from Eq. \eqref{eq:diffuspreding}.}\label{fig:LL}
\end{figure}

The WPG dynamics of Eq. \eqref{eq:mapping00} obviously reproduces the correct effective velocity, defined as $t^{-1}\langle x_i \rangle =  v^{\rm eff}(\theta_i)$ at large times, and since the WPG contains \textit{accurate initial fluctuations} it also reproduces additional hydrodynamic effects such as the diffusive spreading of quasiparticles, namely the quantity $\delta x_i =  t^{-1}\langle (x_i - \langle x_i \rangle )^2 \rangle$ at large times, see Fig. \ref{fig:LL}.  The latter is known to be related to the diffusion constant given by \cite{de_nardis_diffusion_2019,PhysRevB.98.220303}
\begin{equation}\label{eq:diffuspreding}
    \delta x_\theta= \int d\alpha \rho_\alpha f_\alpha |v_\theta^{\rm eff} - v^{\rm eff}_\alpha| \left( \frac{\varphi_{\theta,\alpha}^{\rm dr}}{\rho_{T,\theta}}\right)^2,
\end{equation}
which clearly depends, through the factor $f_\alpha$, on the statistics of the initial fluctuations, which can be chosen to be either quantum or classical. In Appendix~\ref{appendix:sampling}, we sketch the particle sampling procedure used for both classical and Fermi-Dirac statistics and demonstrate that the correct initial fluctuations are present. \\

\section{Generalised fluids (I): integrability breaking via external potentials}
\label{sec: numerics_ext_potentials}
\begin{figure}
  \includegraphics[width=0.48\textwidth]{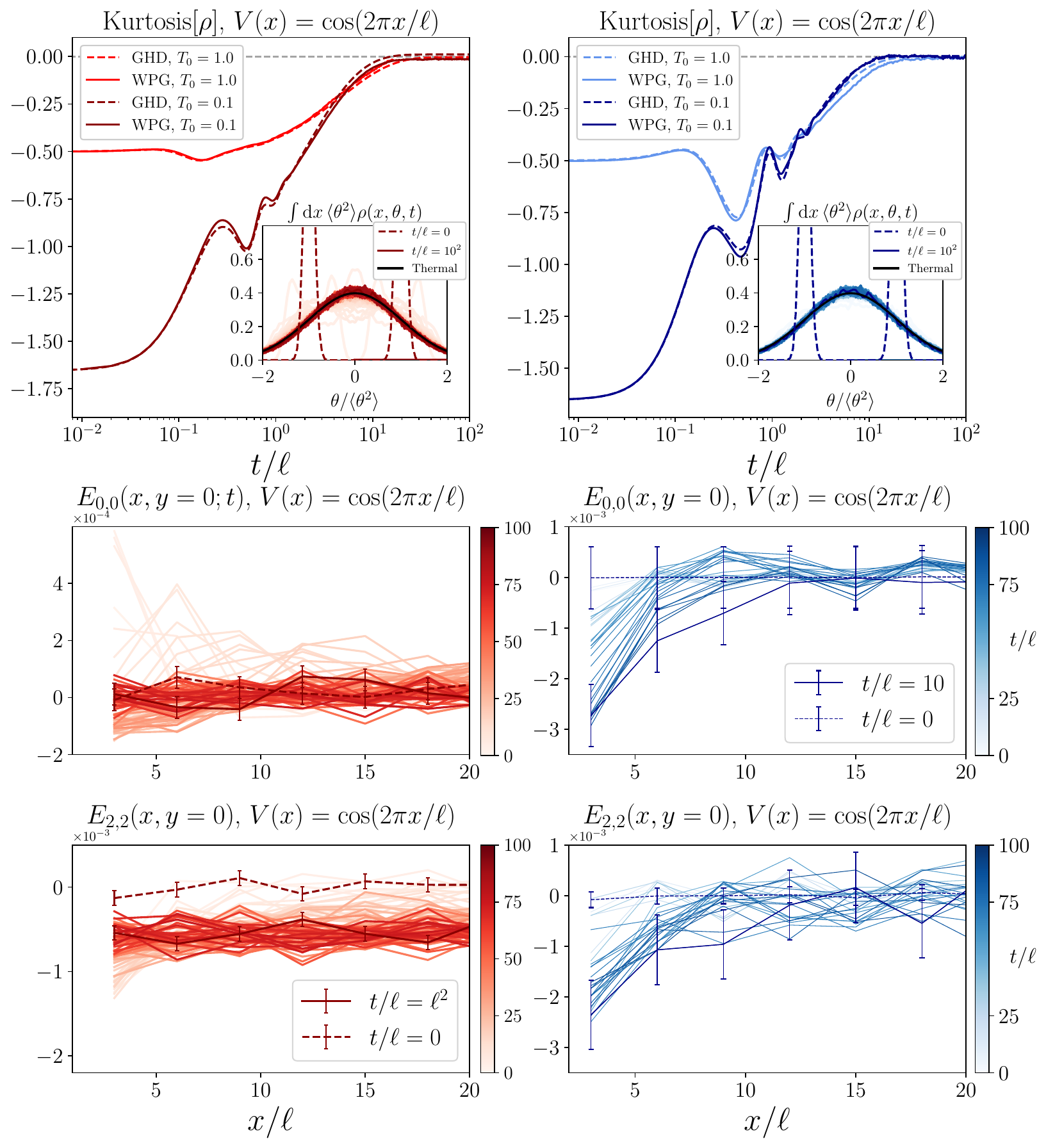}

  \caption{The two columns show respectively the dynamics of classical WPG with  $\mathfrak{a}_{ij} = -1$ (\textit{left, repulsive}) and $\mathfrak{a}_{ij} = +1$ (\textit{right, attractive}). The  initial state is a {Bragg pulse}, given by $\rho = \sum_{s=\pm} \exp{-(\theta -\theta_{s})^2/(2 T_0)}$ with $\theta_{\pm}=\pm 1$. The system is quenched into a cosine potential $V(x) = V_0 \cos( 2\pi x/\ell)$ with $\ell = 10$ and $V_0=1.0$. 
  The top plots show the dynamics of kurtosis of the rapidity distribution as function of rescaled time $t/\ell$ for $T_0=1.0$ and $T_0=0.1$, where solid lines represent the WPG numerics and dashed lines the solution of Navier-Stokes GHD~\eqref{eq:nabla_eq}. The insets show
the dynamics of rescaled rapidity distribution for $T_0=0.01$, where different colors represent different times, according to the colorbars
shown in the plots below. The solid black line shows a featureless Gaussian distribution, while the dashed line represent the
initial distribution.
  The second and third lines show respectively the dynamics of density and energy correlations, defined in~\eqref{eq: def_sp_en_corr}. Different colors represent different increasing times from light to dark, according to the colorbars. 
  }
  \label{fig:CosinePotential}
\end{figure}

Now that we have described how to properly initialize an initial state, we can consider the dynamics, especially in the presence of integrability breaking terms. 
When the dynamics is given entirely by integrable Hamiltonian, the corresponding bare particles positions $X_i(t)$ evolve freely. However, in the presence of an external potential as given by Eq. \eqref{eq:H_inhom}, these trajectories are now modified. The logic is simple: consider the interacting Hamiltonian and map it to bare particles using the bare-to-interacting mapping of Eq. \eqref{eq:mapping00}. While this mapping makes the interacting Hamiltonian $H_0$ trivial, it gives a new effective potential for the bare particles
\begin{equation}
    H_0 + \int \dd{x}V(x)q_{i_{0}}(x)  \to \sum_i \Big[\frac{(v^{\rm bare}(\theta_i))^2}{2} +  V(X_i(x_\cdot))\Big] .
\end{equation}
As the potential is computed at position $X_i(x_\cdot)$ we are required to know the position of the interacting particles $x_i$. The result is a system of interacting particles with an interaction mediated by the external potential.
\\
As an example, let us consider the case of a repulsive hard rods gas in the presence of a quadratic external potential $V(x) = \omega^2 x^2/2$. In this case, the Hamiltonian evolution can be directly mapped to an interacting Hamiltonian for the bare particles as 
\begin{equation}
    H_{\rm HR(a)}[X,\theta]=\sum_i \Big[\frac{\theta_i^2}{2}+\frac{\omega^2 X_i^2}{2}+\frac{a\omega^2}{2}\sum_{j\neq i}|X_i - X_j|\Big]\,.
\end{equation}
The latter corresponds to the Hamiltonian of a self-gravitating gas \cite{Hohl1968, Rybicki1971, Wright1982, Reidl1988, Joyce2010, Teles2011, Roule_2022} confined in a harmonic trap, whose equation of motions are written as 
\begin{equation}
    \dot X_i = \theta_i \quad,\quad \dot \theta_i = -\omega^2 X_i - \frac{a\omega^2}{2} \sum_{i \neq j } {\rm sgn}(X_i - X_j)\,.
\end{equation}
In an analogous way, in generic interacting models the equations of motion can be written via the mapping \eqref{eq:mapping00} but with the bare particles now satisfying the following equations of motions, whose trajectories are shown in Fig. \ref{fig:cartoon}:  
\begin{equation}\label{eq: th_mapped_potential_fp}
\boxed{ \dot X_i = v^{\rm bare}(\theta_i) , \quad \quad 
\dot \theta_i = - (\partial_x V(x) )\bigg|_{x = x_i}.}
\end{equation}
In particular, Eq. \eqref{eq: th_mapped_potential_fp} shows the main complication of this approach: bare particles have a trivial kinetic spreading but they feel a complicated, state dependent, non-local force that is a functional of the interacting particles coordinates.
Hence, it requires transforming between the bare and interacting coordinate systems for each time step and represents the main computational bottleneck of the numerical algorithm. This phenomenon resembles the intrinsic many-body complexity of interacting integrable systems. 
\\
It is important now to stress that, although the single realization of microscopic time evolution is fully deterministic, the average over the initial state ensemble introduces the statistical complexity through the initial state fluctuations, that are transported through the dynamics according to Eq. \eqref{eq:diffusion_full2_qp_trap}.
Hence, this method is not only capable of reproducing the Euler evolution of the gas, but in principle any higher order correction to the hydrodynamics. 
On the other hand, since the $n-$th hydrodynamic corrections produces effects suppressed as $1/\ell^{n-1}$, the expected number of initial states over which perform the average is expected to grow as $\sim\ell^{2(n-1)}$ in order to be able to observe the effect of $n-$th term in the hydrodynamic expansion.
Thus, to show that this method reproduces the correct diffusive order hydrodynamics, we focus on two toy models of WPG: attractive and repulsive hard rods, for which the mapping can be more easily inverted.  
More precisely, in Fig.~\ref{fig:CosinePotential}, we compare diffusive hydrodynamics obtained from WPG simulations with
Navier–Stokes~\eqref{eq:nabla_eq} for the evolution in the periodic potential defined in Eq.~\eqref{eq:cos_pot_def}. As expected \cite{PhysRevResearch.6.023083}, 
for both attractive and repulsive HR the WPG simulations match almost perfectly with the diffusive equation at small times $t/\ell<1$, i.e. when Eq.~\eqref{eq:nabla_eq} is expected to be predictive. This must be considered as a non trivial benchmark that confirms the validity of the numerical method.

\begin{figure*}
  \includegraphics[width=1.0\textwidth]{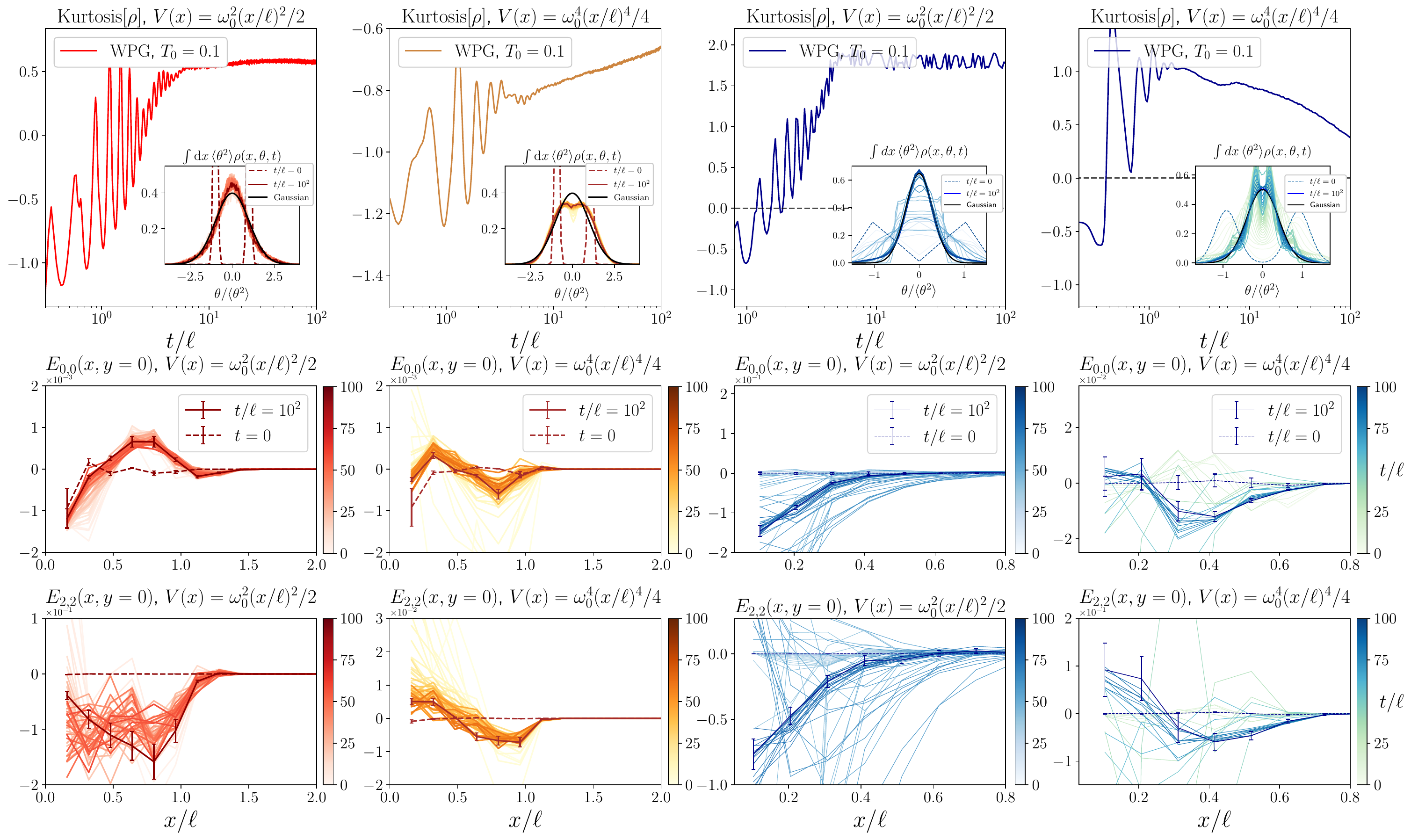}

  \caption{
  The four columns show respectively the dynamics of 
  \textit{repulsive} hard rods evolved in harmonic potential (I), \textit{repulsive} hard rods evolved in quartic potential (II), \textit{attractive} hard rods evolved in harmonic potential (III) and \textit{attractive} hard rods evolved in quartic potential (IV).
  In all the case we set hard rods length $|\mathfrak{a}| = 1 $ and we use $\omega_0 = \sqrt{50}$ for the repulsive case and for the attractive quartic case, while $\omega_0 = \sqrt{5} $ for the attractive harmonic. The initial state is homogeneous in the interval $[-\ell/2,\ell/2]$ with an average number of particles $N=\ell/2=50$ and is prepared in a {Bragg pulse}, given by rapidity distribution $\rho = \sum_{s=\pm} \exp{-(\theta -\theta_{s})^2/(2 T_0)}$, with $\theta = \pm 1$ and $T_0=0.1$.
  The first line shows the WPG dynamics of kurtosis of rapidity distribution against rescaled time $t/\ell$. The insets show the dynamics of rescaled rapidity distribution, where different colors represent different time data according to the colorbars shown in the plots below. The solid black line shows a featureless Gaussian distribution, while the dashed line represent the initial distribution.
  The second and third lines respectively show the dynamics of space and energy correlations defined in~\eqref{eq: def_sp_en_corr}, computed at $y=0$. Different colors represent different increasing times from light to dark, according to the colorbars. Dashed and solid lines are respectively the initial and final time data.
  }\label{fig:HarmonicQuartic-1}
\end{figure*}

\subsection{Simulating trapped integrable systems: thermalisation and long-range correlations}

We shall now use the methods introduced in the previous section to study the dynamics of integrable systems in the presence of external traps. {We consider here, for convenience, the case of attractive and repulsive hard rods}, with scattering shift $\mathfrak{a}_{ij}=\pm1$ and with classical statics. The attractive hard-rod model can be considered as a toy model for the Lieb–Liniger gas at high temperature and therefore its dynamics is directly relevant for experimental realisation. We start by considering particles in a cosine potential
\begin{equation}
\label{eq:cos_pot_def}
    V(x)=V_0 \cos\!\left(2\pi x/\ell\right).
\end{equation}
This potential was also considered in previous works~\cite{PhysRevLett.125.240604,PhysRevResearch.6.023083}, where it was concluded that the system thermalises on, diffusive, time scales proportional to $(\ell/V_0)^2$. Moreover, it was remarked in Ref.~\cite{PhysRevResearch.6.023083} that diffusive hydrodynamics with Navier–Stokes~\eqref{eq:nabla_eq} can capture only the initial Euler-like dynamics and the final thermalisation dynamics, while there is an intermediate time window in which the dynamics are strongly discontinuous (such a region was denoted a turbulent phase) and where hydrodynamics and exact simulation do not fully agree. Such an effect is enhanced by larger values of $\ell$ which prolong the turbulent phase in time.

To benchmark the WPG method a cosine potential is simulated with both the WPG and Navier-Stokes GHD. In this cosine potential case a full simulation with Navier-Stokes GHD is possible, as the state remains smooth during time evolution and no strong singularities in the distribution $\rho_\theta$ are created. We can see that the prediction for the point function agree during the dynamics. In both the repulsive and attractive regimes the distribution of rapidities $\rho(\theta,x)$ converges to the expected thermal distribution (which is also the fixed point of the Navier-Stokes GHD equation Eq. \eqref{eq:nabla_eq}, see \cite{PhysRevResearch.6.023083}), namely a Gaussian (thermal) function,
\begin{equation}
    \rho(\theta,x) \mathrel{\mathop{\longrightarrow}_{t \sim \ell^2}} e^{-\beta\bigl(\theta^{2}+V(x)-\mu\bigr)},
\end{equation}
up to very small deviations. The thermalization predicted by Navier-Stokes GHD can be further interrogated with the WPG method, where higher-point correlations are accessible. For a thermal state these higher-point correlations are expected to become local as the thermal state is approached at the diffusive scale, however an analysis of the two-point correlations suggests a different picture. For simplicity, we focus on two different types of connected correlations, the density-density $\langle N(x) N(y) \rangle^{c}$ and energy-energy $\langle e(x) e(y) \rangle^{c}$, namely 
\begin{equation}
\begin{split}
\label{eq: def_sp_en_corr}
    E_{0,0} &  = \int d\theta d\theta' E_{\theta \theta'} (x,y) = \langle N(x) N(y) \rangle^{c}_{{\rm reg}},\\ 
    E_{2,2} &  = \int d\theta d\theta' 
    \theta^2 \theta'^2  E_{\theta \theta'} (x,y)= \langle e(x) e(y) \rangle^{c}_{\rm reg}, 
\end{split}
\end{equation}
where for regularised correlations $\langle {\cdot}\rangle _{\rm reg}$ we mean that they can be obtained by binning the microscopic system with some mesoscopic space interval, (in this paper we take quite a large one to improve statistics, $dx=3\ell \sim 30$), and we exclude the singular contribution at $x=y$.   
We observe that as the system evolves these correlations develop a finite long-range part (namely, particles correlate over multiple units of $\ell$) and, \textit{even when the one-point function appears thermal, such correlations remain finite, signaling that }the state is not truly thermal\textit{ but is instead a long-range state of the type~\eqref{eq:LRstate}}. Unless stated otherwise, true thermalization would imply a correlation length of the order of the thermal correlation length, which is $O(1)$, whereas here we observe correlation lengths of order $O(10\,\ell)$, which therefore signals that the system has not reached a proper thermal equilibrium even at times of order $\ell^2$. As specified in the previous section, one should therefore expect an even larger time scale of order $\ell^3$ in order, at least in principle, to see two-point functions vanish. Notice that such an effect is clearly due to inter-particle interactions, as in the limit of vanishing interaction one would have only independent trajectories inside the potential.


 We then move to the case of harmonic and quartic traps, namely with $V(x)=\omega^{2}x^{2}/2$ and $V(x)=\omega^{4}x^{4}/4$. To ensure extensive energy, the trap frequency is scaled as $\omega = \omega_0/\ell$ and $\ell=N,$ which corresponds to weak-integrability breaking.  We should first stress that a full simulation with Navier-Stokes GHD in these cases is not possible, as the dynamics become increasingly "turbulent" as function of time, generating strong gradients that are difficult to resolve with finite difference methods.  We first study the repulsive $(\mathfrak{a} = -1)$ and attractive hard rod $(\mathfrak{a} = +1)$ gases prepared in an initial Bragg state, as shown in Figs.~\ref{fig:HarmonicQuartic-1}, and released into the respective traps. For the harmonic trap, the one-point function struggles to thermalise, a phenomenon that was well discussed in previous works~\cite{PhysRevLett.120.164101, PhysRevE.108.064130}. Here we show that, while the distribution of rapidities fails to approach a Gaussian, even the two-point correlations remain finite and large at times of order $\ell^{2}$. Meanwhile, in the quartic case, there are some indications that at times larger than $\ell^{2}$  the one-point function thermalise. This seems consistent with the results of Ref.~\cite{PhysRevE.108.064130} for the quartic potential, where due to strong integrability breaking ($\omega \sim \omega_0$) the one-point function approaches a Gaussian at $t \sim O(N)$. Yet, even in this case correlations functions do not give signs of decaying at these diffusive times, signaling that true thermalisation is not really reached.   \\



\begin{figure*}[t!]
  \includegraphics[width=1.0\textwidth]{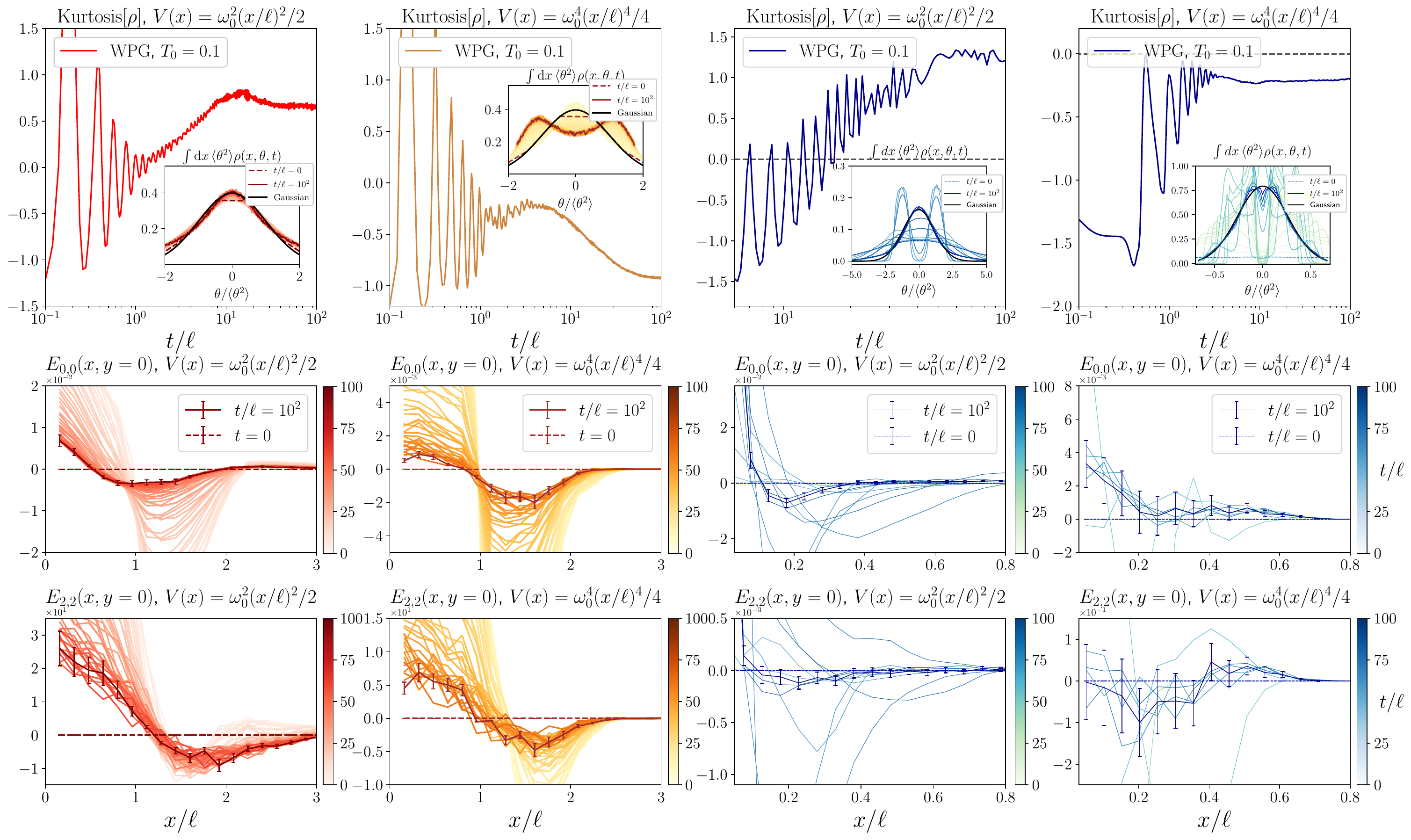} 

  \caption{Same as in Fig. \ref{fig:HarmonicQuartic-1} with the only difference in the initial state preparation:  we initialize the system in a thermal state in the interval $[-\ell,\ell]$ with temperature $T_0=0.01$ and $N=\ell=100$, i.e. homogeneous in space and with Gaussian rapidity distribution.
  Then, at time $t=0$, the particles in $[-\ell/2,\ell/2]$ are removed and we evolve the two {chopped clouds}. }\label{fig:ChopHarmonicQuartic-1}
\end{figure*}

To amplify and elucidate the role of these correlations, we implement a second protocol (Fig.~\ref{fig:ChopHarmonicQuartic-1}): starting with a Gaussian rapidity distribution and uniform spatial density, we remove the particles from the center of the system, creating a Newton's cradle setup, but which can be more efficient realizable experimentally. Strikingly, both harmonic and quartic traps now exhibit a non-Gaussian rapidity distribution and long-range two-point correlations that fail to vanish even at time-scales that are one order of magnitude more than diffusive ones.

Comparisons of the pure trap to the chop protocol make it clear that the two-point correlations are clearly modified between the two settings and that they are strong enough to be detectable in current state-of-the-art experimental platforms. Particularly notable is the very large energy-energy correlator, which clearly supports the idea that there are preferred initial states for investigating these long-range correlators. Therefore, testing different initial states could be used to identify promising experimental protocols that could in turn be realized in cold atom experiments to investigate the emergence of these long-range correlations.

\section{Generalised fluids (II): integrability breaking via two-body potential}
\label{sec: numerics_two_body_pot}


\begin{figure}[h!]
    \centering
  \includegraphics[width=0.49\textwidth]{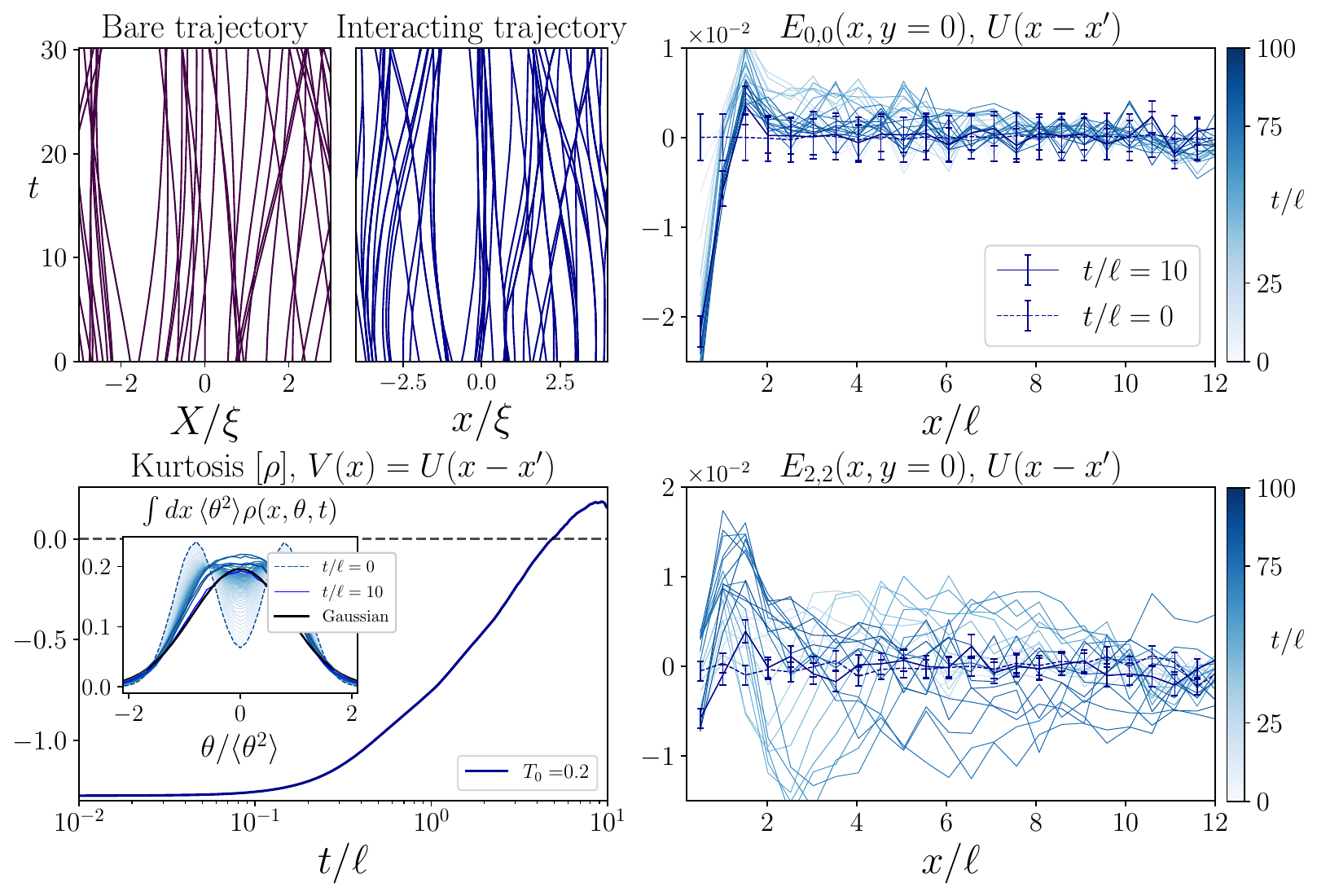}
   \caption{Dynamics of classical attractive WPG with $\mathfrak{a}_{ij} = 1$ and long-range potential \eqref{eq:longrange_potential} with $\ell = 20$. The initial state is a Bragg split state with $\theta = \pm 1$ and $T_0=0.01$. The top left panel shows an example of microscopic trajectories for both interacting and bare coordinates. Bottom left figure shows the dynamics of kurtosis of rapidity distribution. The inset shows the rapidity distribution at different times, compared with the expected Gaussian thermal shape (black dashed line).  
   The two figures on the right show the dynamics of density and energy correlations defined in~\eqref{eq: def_sp_en_corr}. Different colors are associated with different increasing times from light to dark, as explained by the colorbar. The dashed and solid lines are respectively the initial and final time data. 
   }
   \label{fig:ddi_attractive}
\end{figure}

The approach introduced in the previous sections allows us to tackle more complex integrability-breaking potentials.  
In this section, we consider integrable systems perturbed by two-body interactions, a scenario motivated by recent cold-atom experiments.  
For example, in \cite{Tang2018}, a gas of ultracold bosonic dysprosium atoms is tightly confined in one-dimensional tubes formed by a two-dimensional optical lattice.  
These atoms exhibit dipole–dipole interactions (DDI) roughly 100 times stronger than those of rubidium.  
Whereas ultracold bosons usually interact locally and are well described by the integrable Lieb–Liniger model, the DDI introduces a long-range, two-body, integrability-breaking interaction that drives late-time thermalization of the gas, namely the system is described by the Hamiltonian 
\begin{equation}
    H = H_{\rm Lieb-Liniger} + \sum_i V(x_i) +  \sum_{i<j} U(x_i - x_j).
\end{equation} 
As shown in \cite{Biagetti2025}, the hydrodynamic evolution of such a quasi-integrable gas is governed by a hierarchy of nonlinear integro-differential equations, analogous to the BBGKY hierarchy.  
Under the assumptions of weak interactions and a large hydrodynamic scale $\ell$, this hierarchy can be closed, but its numerical solution remains computationally prohibitive.  
We now illustrate how the numerical method developed here can efficiently simulate the hydrodynamics of these systems.

Consider the long-range potential
\begin{equation}
\label{eq:longrange_potential}
U(x-x') \;=\; \frac{U_0}{\ell}
\biggl[1 + \bigl(\tfrac{|x - x'|}{\ell}\bigr)^{3}\biggr]^{-1},
\end{equation}
written in interacting coordinates and evolved via the bare-particle mapping.  
Our wave-packet gas (WPG) algorithm applies directly, requiring only the coordinate transformations between bare and interacting frames.  
In the presence of an external trap $V(x)$, the equations of motion become
\begin{align}
\dot X_i &= v^{\rm bare}(\theta_i),\\
\label{eq:LR_th_bare}
\dot \theta_i &= -\bigl.\partial_x V(x)\bigr|_{x=x_i}
\;-\;\sum_{k\neq i}\bigl.\partial_x U(x)\bigr|_{x=x_i - x_k}.
\end{align}
In bare coordinates, the two-body integrability-breaking potential acquires dependence on the interacting-particle positions, yielding a $1/\ell$ expansion of many-body interactions in which the $n$-body term scales as $1/\ell^{n-2}$.

As an illustrative example, Fig.~\ref{fig:ddi_attractive} shows the dynamics of an attractive WPG with $\mathfrak{a}_{ij}=c$, interacting via the potential \eqref{eq:longrange_potential}.  
This toy model approximates the high-temperature limit of a DDI-interacting Lieb–Liniger gas.  
Starting from a Bragg-pulse initial state, we track the density and connected correlations.  
The kurtosis of the rapidity distribution relaxes rapidly on a time scale set by the integrability-breaking strength.  
After this fast relaxation, the density reaches a quasi-steady plateau and then drifts slowly toward the thermal distribution.  
Simultaneously, intermediate-time correlations build up and relax to a correlation length of order $\ell$.  
Since the true thermal state exhibits correlations of the same range, Fig.~\ref{fig:ddi_attractive} confirms that the system indeed fully thermalizes in this case.

\section{Conclusion}
\label{sec: conclusions}

In this work, we have shown that the semiclassical wave-packet formulation of GHD \cite{PhysRevLett.132.251602,10.21468/SciPostPhys.13.3.072,Bonnemain2025,2503.08018,Hubner2023} provides a transparent and efficient algorithm for numerically evolving both classical and quantum integrable systems, including all orders of fluctuating hydrodynamics.  
Importantly, our method can incorporate arbitrary integrability-breaking terms.  
On one hand, it overcomes the challenge of solving the full hydrodynamic hierarchy for a quasi-integrable model, yielding all-order hydrodynamics efficiently; on the other, it embeds generic perturbations—many-body interactions, particle loss, or impurity scattering—simply by adding the corresponding forces to the bare-particle dynamics.

Motivated by recent cold-atom experiments, we applied the WPG algorithm to two paradigmatic scenarios: an external periodic potential (Eq.~\eqref{eq:cos_pot_def}) and a long-range dipolar two-body interaction.  
In the periodic trap, at times $O((\ell/V_0)^2)$ the density relaxes to a nearly Gaussian form, consistent with previous findings \cite{PhysRevResearch.6.023083}, however, correlations reveal persistent far from thermal long-range terms that are built-up at intermediate time-scales and that fail to dissipate at large times.  
According to Eq.~\eqref{eq: diff_correction_correlations}, two-point functions are expected to relax on the much longer scale $\ell^3$. On the other hand,   
for polynomial traps, in agreement with Refs.~\cite{PhysRevLett.120.164101,PhysRevE.108.064130}, the rapidity distribution remains non-Gaussian and long-range correlations persist for even longer times, indicating that true thermalization is still yet to be reached. The final reason for such a failure of dissipation of correlations is left for future works, even if the bare-to-interacting mapping shown here seems to point to the fact that the integrable anyway the weakly mixing nature or trajectories of the associated bare particles may play an important role.  
Moreover, varying initial states strongly affect the generation of long-range correlations, suggesting that tailored initial conditions could enable experimental protocols to probe correlation dynamics in cold-atom setups.

Our findings raise new questions about the role of long-range many-body correlations in the relaxation of quasi-integrable systems, particularly regarding the absence of thermalization in harmonic confinement.  
They also open exciting directions for studying nearly integrable systems with diverse integrability-breaking mechanisms, potentially modeling realistic features of cold-atom experiments.

\begin{acknowledgments}
A.U., L.B, J.K and J.D.N. are funded by the ERC Starting Grant 101042293 (HEPIQ) and the ANR-22-CPJ1-0021-01.  This work was granted access to the HPC resources of IDRIS under the allocation AD010613967R2.   We thank Alvise Bastianello, Frederik Møller, Jorg Schmiedmayer, Benjamin Doyon, Romain Vasseur, Sarang Gopalakrishnan, Takato Yoshimura, Friedrich Hübner and Herbert Spohn for valuable feedbacks. 
\end{acknowledgments}
\appendix

\section{Dynamics induced by bare-to-interacting mapping and GHD}\label{app:ghdmicro}

We present a microscopic derivation of the Euler GHD equation for the evolution of quasiparticle density
\begin{equation}
    \rho(x,\theta) \equiv   \sum_i \delta(x-x_i)\delta(\theta - \theta_i) \quad . \quad
\end{equation}
Here, the quasiparticle position $x_i$ evolves according to the mapping of Eq.~\eqref{eq:mapping00}, where the bare particles' positions $X_i$ evolve with the bare velocities $\dot{X}_i = v^{\rm bare}(\theta_i)$. We note that the motion of the quasiparticles is given by the dressed velocity $\dot x_i = v^{\rm{eff}}_{[\rho(x_i,\cdot)]}(\theta_i)$.

Taking the time derivative of Eq.~\ref{eq:mapping00} we get
\begin{equation}
   v^{\rm bare}(\theta_i) = \dot X_i = \dot x_i + \sum_{j\neq i}  \mathfrak{a}_{ij} \delta(x_i - x_j) (\dot x_i - \dot x_j) \, ,
\label{eq: map derivative}
\end{equation}
where $\dot x_i$ is a function of all the particles' coordinates defined by Eq.~\eqref{eq: map derivative}. Taking the time derivative of the density of particles, we get 
\begin{align}
    \partial_t \rho(x,\theta) &= \sum_i \partial_t(\delta(x-x_i)\delta(\theta - \theta_i)) 
    \\
    &=-\sum_i \partial_{x}(\delta(x-x_i)\delta(\theta - \theta_i)) v^{\rm{eff}}_{[\rho(x_i,\cdot)]}( \theta_i) \notag 
    \\
    &= -\partial_x \bigg(v^{\rm{eff}}_{[\rho(x,\cdot)]}( \theta)\sum_i(\delta(x-x_i)\delta(\theta - \theta_i)) \bigg), \notag
\end{align}
where we used the following delta function property 
$\partial_x \delta(x-y) f(y) =  \partial_x (\delta(x-y) f(x))
$. We therefore obtain
\begin{equation}\label{eq:MGHD1}
     \partial_t \rho(x,\theta) + \partial_x \big(v^{\rm{eff}}_{[\rho(x,\cdot)]}( \theta)\rho(x,\theta)\big) =0
\end{equation}
Also, we have introduced the effective velocity
\begin{align}
    v_{[\rho(x_i, \cdot)]}^{\rm eff}(\theta_i) &= \dot x(x_i, \theta_i) \notag \\
&=  v^{\rm bare}(\theta_i)\\ &+   \sum_{j\neq i} \mathfrak{a}_{ij} \delta(x_i - x_j) (\dot x(x_i, \theta_i) - \dot x(x_j, \theta_j)).\notag
\end{align}
By replacing the sum with a coarsegrained integral {\it i.e.}, $\sum_{i, j}.. \to \int dx' \int d\theta' \rho(x', \theta') .. $ we get
\begin{align} \label{eq:MGHD2}
& v_{[\rho(x_i,\cdot)]}^{\rm eff}(\theta_i) = v^{\rm bare}(\theta_i) \\&-    \int d{\theta'}  \mathfrak{a}(\theta_i - \theta') \rho(x_i,\theta')   (v_{[\rho(x_i,\cdot)]}^{\rm eff}(\theta_i) - v_{[\rho(x_i,\cdot)]}^{\rm eff}(\theta')). \notag 
\end{align}
Equations \eqref{eq:MGHD1} and \eqref{eq:MGHD2} are valid even at the microscopic level, namely, they are exact for any single trajectory of particle coordinates. 

\section{Sampling initial state for the WPG}\label{appendix:sampling}
Below we sketch our procedures for sampling particle position and rapidity such that the initial state contain the correct fluctuations, which can be checked against known initial correlation functions. These procedures are carried out for homogeneous and inhomogeneous states, separately considering both classical and quantum (Fermi-Dirac) systems.

\subsection{Sampling a classical homogeneous stationary state: } 
We here describe how to sample the initial particle positions and rapidities of an initial stationary homogeneous state, described by a given density of rapidities $\rho_\theta$  

\begin{enumerate}

    \item Identify the target density for a particle species, $\rho = \int d\theta \rho(\theta)$, along with a fixed length $L$ for the interacting model, such as a hard rod gas (HR). From this given target density, the bare particle density $\rho^{\rm bare}$ is determined as \begin{eqnarray}
        \rho_{\rm bare}(\theta) = \frac{\rho(\theta)}{2\pi \rho_T (\theta)} =  \frac{n(\theta)}{2\pi}  
    \end{eqnarray}
    with $\rho_T(\theta)$ being the total density of states.
    \item Initialize the bare positions $X_i$:  draw a distance $r_i$ from an exponential distribution weighted by the integrated bare density,
    \begin{eqnarray}
        P^{(s)}(r) \sim e^{-r\int d\theta \, \rho_{\rm bare}(\theta)},
    \end{eqnarray} 
    a procedure known as a Poisson point process. A bare particle is then placed a distance $r_i$ away from the previously placed particle, or the origin. The Poisson point process generates an initially uncorrelated state in the bare coordinates.
    \item Assign a momentum to the newly placed bare particle by sampling the distribution of momenta $\rho(\theta)$.
    \item Transform from bare coordinates to interacting ones by inverting Eq.~\eqref{eq:mapping00}. Break the loop if the transformed position of the new particle is outside the system size $L$; if not, return to step 2.
\end{enumerate}

\begin{figure}[h!]
    \centering
    \includegraphics[width=1.0\linewidth]{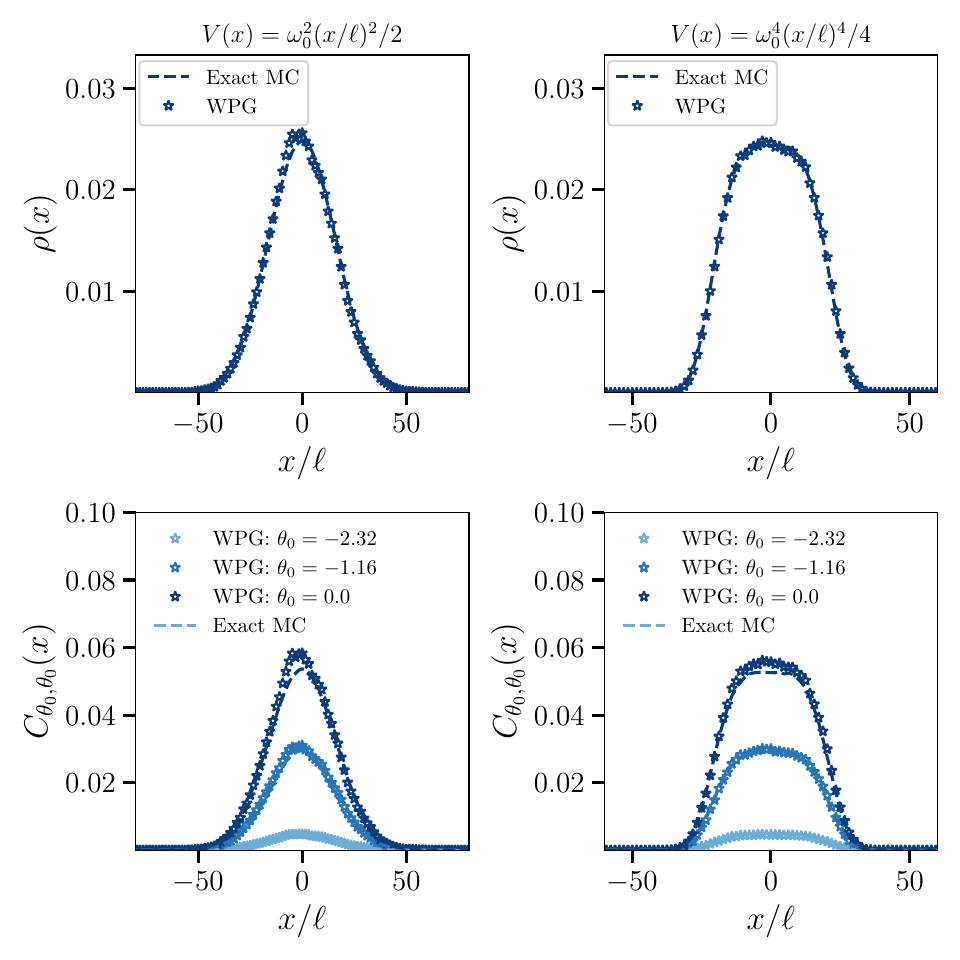}
    \caption{The figure shows the density profile $\rho(x)$ (top row) and the susceptibility (bottom row)  {\it i.e.} $ \langle \delta \rho(x',\theta_0) \delta \rho(x,\theta_0) \rangle_c = C_{\theta_0, \theta_0}(x)\delta(x-x')$ for WPG gas with the scattering kernel of the Toda model at equilibrium with $\beta = 1.0$. We confine the WPG in two different traps: harmonic (left column) and quartic (right column) with $\ell = 16$ and $\omega_0 = 1/16$ for $N=16$ particles.  These profiles (star symbols) are obtained using the sampling procedure for classical inhomogenous systems described in the text, and the dashed lines represent the MC simulation of the Toda model. For computing correlation, we set $dx = 5$ and average over $10^5$ samples.}
    \label{fig:WPG-Toda}
\end{figure}

\subsection{Sampling a classical inhomogeneous equilibrium state: }
We here describe how to correctly sample the initial particle positions and rapidities of an initial stationary inhhomogeneous state, as for example the stationary state of a classical system inside an external potential $V(x)$

\begin{enumerate}
    \item To obtain thermal density profile at a temperature $\beta$ when the system is confined in a trap $V(x)$. We use Monte-Carlo (MC) simulations, with an energy function 
    \begin{align}
        H(\{\theta_i, X_i\}) = \sum_i \frac{\theta_i^2}{2} + \sum_iV(x_i),
    \end{align}
    where $x_i$ is a function of $\{\theta_j,X_j\}$ and obtained by inverting Eq.~\eqref{eq:mapping00}.
    \item In the MC procedure, we randomly choose a bare particle and perturb its rapidity and position $(\theta_j, X_j) \to (\theta_j', X_j')$. Under this perturbation, we compute the energy difference $\delta H=H'-H$ where $H$ and $H'$ are the energies of the old and new configurations, respectively.
    \item Sample a random variable $r$ from uniform distribution between $r \in [0,1]$, if $r<\exp(-\beta \delta H)$ accept the configuration otherwise reject. 
    \item Save the interacting particle configuration $\{\theta_i, x_i\}$ after $N$ such steps.
    \item Bin the particles in the resulting interacting phase space with discretisation $d\theta dx = 2 \pi$. \\

    This protocol is verified in Fig. \eqref{fig:WPG-Toda} which shows the density profile and two-point correlations for a Toda chain at thermal equilibrium inside a confining potential. We compare the result obtained from WPG and from exact Monte Carlo (MC) simulations, obtaining excellent agreement. 
\end{enumerate}

\subsection{Sampling a quantum (fermionic) homogenous stationary state: } 
We here introduce a procedure analogous to the classical homogenous case but done to incorporate fermionic quantum statistics.  
\begin{enumerate}
    \item Introduce a trapping potential so that outside of a fixed region there will be no particles.
    \item Here, the initial input will be the occupation function, $n(\theta,X) = n(\theta)$, determined from TBA and determines the bare particle density as
    \begin{eqnarray}
        \rho^{\rm bare}(\theta)  = \frac{n(\theta)}{2\pi} .
    \end{eqnarray}
    Chop the bare phase space into a set of bins subject to the quantization condition $d\theta dX = 2\pi$.
    \item Sample a random number $p_{ij}$, whenever $p_{ij} < n(\theta_i,X_j)$ a particle is inserted into the corresponding phase space bin.
    \item Transform from bare coordinates to interacting ones by inverting Eq.~\eqref{eq:free_to_int_rWPG}. Re-bin the particles in the resulting interacting phase space according to the same discretization $d\theta dX = d\theta dx = 2 \pi$. \\

    This protocol is verified in Fig. \eqref{fig:LL} which shows how this initialization of the state not only guarantees that the effective velocities are correct but also that its fluctuations, giving the diffusion spreadings, are. 
\end{enumerate}

\subsection{Sampling a quantum (fermionic) inhomogeneous stationary state: }
\begin{figure}[h!]
    \centering
    \includegraphics[width=1.0\linewidth]{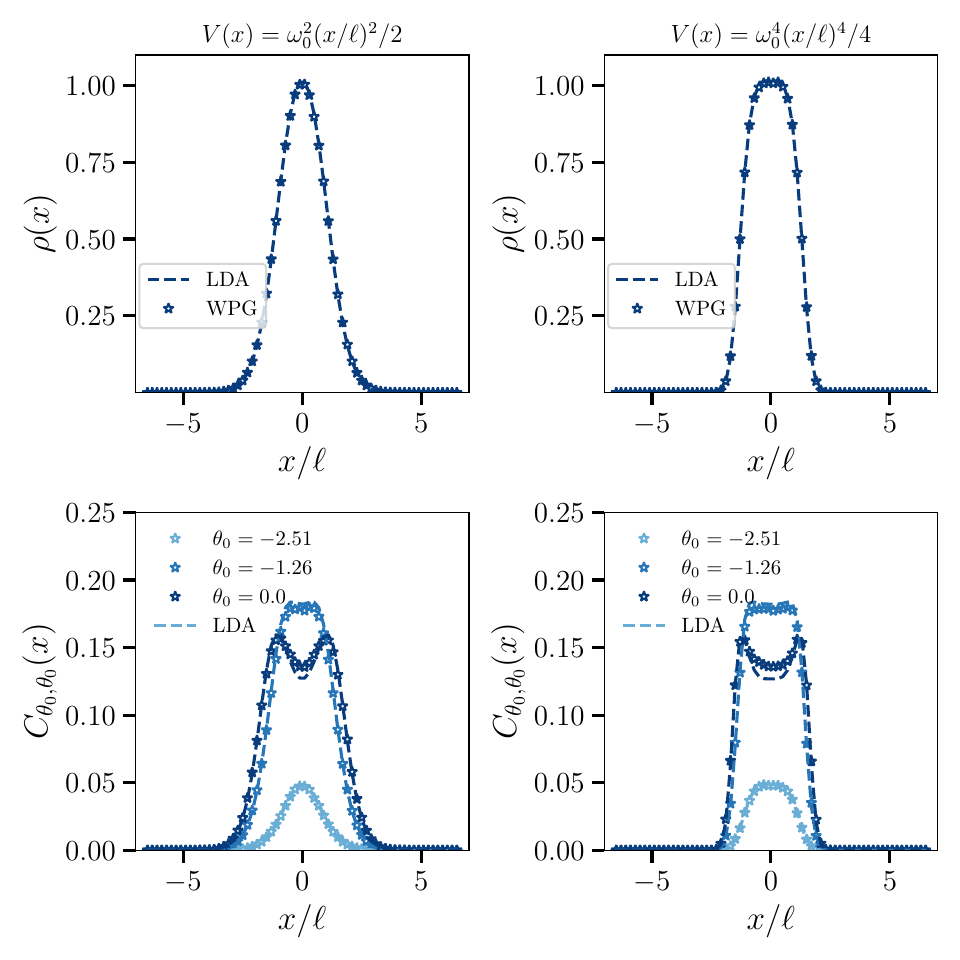}
    \caption{The figure shows the density profile $\rho(x)$ (top row) and the density-density  susceptbility (bottom row)  {\it i.e.} $ \langle \delta \rho(x',\theta_0) \delta \rho(x,\theta_0) \rangle_c = C_{\theta_0, \theta_0}(x)\delta(x-x')$ for WPG gas with the scattering kernel of the Lieb-Liniger model with $c=1$ at equilibrium with $\beta = 1.0$ and $\mu=1.0$. We confine the WPG in two different traps: harmonic (left column) and quartic (right column) with $\ell = 100$ and $\omega_0 = 1.0$.  These profiles (star symbols) are obtained using the sampling procedure for quantum inhomogenous systems described in the text, with $dX = 5$ and the average is over $10^5$ samples. The local density approximation (LDA) prediction is obtained by considering a gas at equilibrium with $\mu \to \mu - V(x)$.}
    \label{fig:WPG-initial-conditions-check}
\end{figure}

Similarly to the classical case, we consider a stationary state of a system with fermionic statistics in an external potential $V(x)$.
\begin{enumerate}
    \item Compute the occupation function $n(\theta,x)$ using the TBA with the local-density-approximation (LDA) chemical potential $\mu \to \mu-V(x)$. The trap $V(x)$ is such that it naturally confines the particles.
    \item Discretize the phase space into a set of bins with the quantization condition $d\theta dX = 2 \pi$. 
    \item Compute the bare particle density at any bin located at $(\theta_i, X_j)$ using
    \begin{eqnarray}\label{density-mapping}
        \rho^{\rm bare}(\theta_i, X_j) = n(\theta_i,x)
    \end{eqnarray}
    where $x$ is a function of $(\theta_i, X_j)$ and is obtained by numerically solving coarsegrained Eq.~\eqref{eq:mapping00} which is given by
    \begin{align}\label{coarse-mapping}
        &x = X_j- \frac{1}{2} \iint dx' d\theta' \rho(\theta',x')\mathfrak{a}(\theta_i-\theta')\mathrm{sign}(x-x').
    \end{align}
    Here $\rho(\theta,x)$ is the phase space density. Note that we use interpolated $n(\theta,x)$ in Eq.~\eqref{density-mapping}. 
    \item For each $i$ and $j$, sample a random number $p_{ij}$ from uniform distribution between $p_{i,j} \in [0,1]$, whenever $p_{ij} < \rho^{\rm bare}(\theta_i, X_j)$, a particle is inserted into the corresponding phase space bin.
    \item Transform from bare coordinates to interacting ones by inverting Eq.~\eqref{eq:mapping00}. Re-bin the particles in the resulting interacting phase space according to the same discretization $d\theta dX = d\theta dx = 2 \pi$. \\
    
    This protocol is verified in Fig. \eqref{fig:WPG-initial-conditions-check} which shows the density profile and two-point correlations for the Lieb-Liniger model at thermal equilibrium in the presence of an external potential. We compare the result obtained from WPG and from local density approximation (LDA), obtaining excellent agreement. 
    
\end{enumerate}

\section{Diffusive hydrodynamics from long-range correlations in inhomogeneous Hamiltonians}
\label{sec: app_diffusive_inhomogeneous}
In this section we explicitly derive the  Eq.~\eqref{eq:diffusion_full1_qp_trap} and~\eqref{eq:diffusion_full2_qp_trap} introduced in the main text. Then, we show that, in the small time limit $t\to0^+$, they reproduce the same diffusive equation shown in
\cite{Durnin2021}, therefore confirming the general statement \textit{that Navier-Stokes equation only applies in the limit of short times}. 
\subsection{Derivation of Eq.~\eqref{eq:diffusion_full1_qp_trap} and~\eqref{eq:diffusion_full2_qp_trap}}
We firstly consider the equation for the quasiparticles density $\rho$. In the same spirit as \cite{hubner2024diffusive}, we aim to write the diffusive correction to the Euler dynamics using the hydrodynamic expansion of currents. In the case of inhomogeneous forcing fields, we must consider the double projection of both currents proportional to the effective velocity and rapidity
\begin{eqnarray}
\label{eq: diffusion_full1_qp_trap_app}
    \partial_t \rho_\theta
    &&+ \partial_x (v^{\text{eff}}_\theta \rho_\theta)
    + \partial_\theta (a^{\text{eff}}_\theta \rho_\theta)=
    \nonumber\\
    &&= \frac{1}{2}\,\partial_x \!\bigl( H_\theta^{\alpha\beta} E^{\text{sym}}_{\alpha\beta} \bigr)+ \frac{1}{2}\,\partial_\theta \!\bigl( H_\theta^{\mathfrak{f},\alpha\beta} E^{\text{sym}}_{\alpha\beta} \bigr)\,.
\end{eqnarray}
In particular we have defined (see Appendix's section~\ref{sec: derivation_H_matrix_app} for a derivation)
\begin{eqnarray} 
\label{eq: H_matr_app}
H_{\theta}^{\alpha\beta}&\equiv&\frac{\delta^2 v^{\rm eff}_{\theta}}{\delta\rho_{\alpha}\delta\rho_{\beta}}
\\
\nonumber
&=&
R^{-t}_{\theta\gamma}
    \Delta v^{\text{eff}}_{\mu,\gamma}\tfrac{\varphi^{\text{dr}}_{\gamma\mu}}{\rho_{T,\gamma}}
(R^{t}_{\mu\alpha} R^{t}_{\gamma\beta}+R^{t}_{\mu\beta} R^{t}_{\gamma\alpha})\,,
\\
\label{eq: Hf_matr_app}
H_{\theta}^{\mathfrak{f},\alpha\beta}&\equiv&\frac{\delta^2 a^{\rm eff}_{\theta}}{\delta\rho_{\alpha}\delta\rho_{\beta}}
\\\nonumber&=& R^{-t}_{\theta\gamma}
    \Delta a^{\text{eff}}_{\mu,\gamma}\tfrac{\varphi^{\text{dr}}_{\gamma\mu}}{\rho_{T,\gamma}}
(R^{t}_{\mu\alpha} R^{t}_{\gamma\beta}+R^{t}_{\mu\beta} R^{t}_{\gamma\alpha})\,,
\end{eqnarray}
with the definitions $\Delta v^{\text{eff}}_{\mu,\gamma}\equiv v^{\text{eff}}_{\mu}-v^{\text{eff}}_{\gamma}$ and $\Delta a^{\text{eff}}_{\mu,\gamma}\equiv a^{\text{eff}}_{\mu}-a^{\text{eff}}_{\gamma}$
and where we considered the point splitting regularization $E^{\rm sym}_{\alpha,\beta}=(E_{\alpha,\beta}(x^+,x^-)+E_{\alpha,\beta}(x^-,x^+))/2$. Also, in the whole section we use Einstein index notation, assuming integration over repeated indexes and more precisely considering unintegrated any indexes appearing on both l.h.s and r.h.s.\\
The evolution equation for the connected two-point function reads
\begin{multline}
    \partial_t \langle \rho_{\theta}(x)\rho_{\theta'}(x')\rangle^c + \Big[\partial_x\big( A_{\theta}^{\gamma}\langle \rho_{\gamma}(x)\rho_{\theta'}(x')\rangle^c \big)+\\+ \partial_{\theta}\big( A_{\theta}^{\mathfrak{f},\gamma}\langle \rho_{\gamma}(x)\rho_{\theta'}(x')\rangle^c \big)+(x,\theta\leftrightarrow x',\theta')\Big]=0\,,
\end{multline}
together with the matrices
\begin{equation}
    A_{\theta}^{\theta'} = R^{-t}_{\theta,\gamma}v^{\rm eff}_{\gamma}R^t_{\gamma,\theta'}\quad,\quad A_{\theta}^{\mathfrak{f},\theta'} = R^{-t}_{\theta,\gamma}a^{\rm eff}_{\gamma}R^t_{\gamma,\theta'}\,.
\end{equation}
In order to derive an equation for the regular part of the connected correlator we must use the decomposition
\begin{equation}
    \langle \rho_{\theta}(x)\rho_{\theta'}(x')\rangle^c = \delta(x-x')C_{\theta,\theta'}(x)+E_{\theta,\theta'}(x,x'),
\end{equation}
being $C_{\theta,\theta'}=R^{-t}_{\theta,\gamma}f_{\gamma}\rho_{\gamma}R^{-1}_{\gamma,\theta'}$ the system equilibrium susceptibility matrix.
Hence, we can write 
\begin{equation}
\label{eq: evo_E_implicit_app}
\begin{split}
 &\partial_t E_{\theta,\theta'}(x,x') + \Big[\partial_x\big( A_{\theta}^{\gamma}E_{\gamma,\theta'}(x,x') \big)+\\&+ \partial_{\theta}\big( A_{\theta}^{\mathfrak{f},\gamma}E_{\gamma,\theta'}(x,x') \big)+(x,\theta\leftrightarrow x',\theta')\Big]
    \\&=\delta(x-x')\Big(\partial_t C_{\theta,\theta'} + \partial_x B_{\theta,\theta'}+(\partial_{\theta}+\partial_{\theta'})B^{\mathfrak{f}}_{\theta,\theta'}\Big)\,,
\end{split}
\end{equation}
with the matrices 
\begin{eqnarray}
    B_{\theta,\theta'}&=&A_{\theta}^{\gamma}C_{\gamma,\theta'}=R^{-t}_{\theta,\gamma}v^{\rm eff}_{\gamma}f_{\gamma}\rho_{\gamma}R^{-1}_{\gamma,\theta'}\,,\\
    \label{eq: def_B_f_app}
    B^{\mathfrak{f}}_{\theta,\theta'}&=&A_{\theta}^{\mathfrak{f},\gamma}C_{\gamma,\theta'}=R^{-t}_{\theta,\gamma}a^{\rm eff}_{\gamma}f_{\gamma}\rho_{\gamma}R^{-1}_{\gamma,\theta'}\,.
\end{eqnarray}
We now proceed by computing explicitly the term proportional to the Dirac delta function in terms of derivatives of the quasiparticles density
\begin{equation}
\begin{split}
    \label{eq: der_Cmatr_app}
    &\partial_t C_{\theta,\theta'} = \frac{\delta C_{\theta,\theta'}}{\delta\rho_{\gamma}}\partial_{t}\rho_{\gamma}=\\&=-\frac{\delta (R^{-t}_{\theta,\gamma}f_{\gamma}\rho_{\gamma} R^{-1}_{\gamma,\theta'})}{\delta \rho_{\zeta}} \big[\partial_x(v^{\rm eff}_{\zeta}\rho_{\zeta})+\partial_\zeta(a^{\rm eff}_{\zeta}\rho_{\zeta})\big]\,,
\end{split}
\end{equation}
where we use the leading order of~\eqref{eq: diffusion_full1_qp_trap_app}.
In addition, we also compute the following derivatives (see Appendix's section~\ref{eq: derivation_M_matrix_app} for a derivation)
\begin{eqnarray}
\label{eq: der_Bmatr_app}
    \partial_x B_{\theta,\theta'}&=& R^{-t}_{\theta,\gamma}f_{\gamma}\partial_{x}(v^{\rm eff}_{\gamma}\rho_{\gamma})R^{-t}_{\gamma,\theta'}+\\\nonumber&+&\frac{\delta (R^{-t}_{\theta,\gamma}f_{\gamma} R^{-1}_{\gamma,\theta'})}{\delta \rho_{\zeta}}v^{\rm eff}_{\gamma}\rho_{\gamma} \partial_x\rho_{\zeta}\,,
    \\
    \label{eq: der_Bmatr_f_app}(\partial_{\theta}+\partial_{\theta'})B^{\mathfrak{f}}_{\gamma,\theta'}
    &=&R^{-t}_{\theta,\gamma}f_{\gamma}\partial_{\gamma}(a^{\rm eff}_{\gamma}\rho_{\gamma})R^{-t}_{\gamma,\theta'}
    \\\nonumber&+&\frac{\delta (R^{-t}_{\theta,\gamma}f_{\gamma} R^{-1}_{\gamma,\theta'})}{\delta \rho_{\zeta}}a^{\rm eff}_{\gamma}\rho_{\gamma} \partial_\zeta\rho_{\zeta}\,.
\end{eqnarray}
Summing the results~\eqref{eq: der_Cmatr_app},~\eqref{eq: der_Bmatr_app} and \eqref{eq: der_Bmatr_f_app} we find
\begin{eqnarray}
\label{eq: sum_der_matr_app}
    &&\nonumber\partial_t C_{\theta,\theta'} + \partial_x B_{\theta,\theta'}+(\partial_{\theta}+\partial_{\theta'})B^{\mathfrak{f}}_{\theta,\theta'}=
    \\&&
    =\frac{\delta (R^{-t}_{\theta,\gamma}f_{\gamma} R^{-1}_{\gamma,\theta'})}{\delta \rho_{\zeta}}\rho_{\gamma} R^{-t}_{\zeta,\xi}\Delta v^{\rm eff}_{\gamma,\xi}R^{t}_{\xi,\alpha}\partial_x\rho_{\alpha}
    \\\nonumber&&
    +\frac{\delta (R^{-t}_{\theta,\gamma}f_{\gamma} R^{-1}_{\gamma,\theta'})}{\delta \rho_{\zeta}}\rho_{\gamma} R^{-t}_{\zeta,\xi}\Delta a^{\rm eff}_{\gamma,\xi}R^{t}_{\xi,\alpha}\partial_\alpha\rho_{\alpha}\,,
\end{eqnarray}
where we used the relation
\begin{multline}
    \partial_x(v^{\rm eff}_{\theta}\rho_{\theta})+\partial_\theta(a^{\rm eff}_{\theta}\rho_{\theta})=
    \\=R^{-t}_{\theta,\gamma}v^{\rm eff}_{\gamma}R^{t}_{\gamma,\zeta}\partial_x\rho_{\zeta}+R^{-t}_{\theta,\gamma}a^{\rm eff}_{\gamma}R^{t}_{\gamma,\zeta}\partial_\zeta\rho_{\zeta}\,.
\end{multline}
Using the result~\eqref{eq: sum_der_matr_app} into Eq.~\eqref{eq: evo_E_implicit_app}, we finally find the explicit form for the r.h.s 

\begin{eqnarray}
\label{eq: diffusion_full2_qp_trap_app}
    \partial_t E_{\theta\theta'}&&(x,x')
    + \partial_x (A_\theta^{\;\alpha} E_{\alpha\theta'})
    + \partial_{x'} (A_{\theta'}^{\;\alpha} E_{\theta\alpha})+
    \nonumber\\
    &&+ \partial_\theta (A_\theta^{\mathfrak{f},\gamma} E_{\gamma\theta'})
    + \partial_{\theta'} (A_{\theta'}^{\mathfrak{f},\gamma} E_{\theta\gamma})=
    \nonumber\\
    &&=
    \delta(x-x')\bigl( M_{\theta\theta'}^{\;\alpha} \partial_x \rho_\alpha
    + M_{\theta\theta'}^{\mathfrak{f},\alpha} \partial_\alpha \rho_\alpha \bigr)\,
\end{eqnarray}
together with the definition
\begin{eqnarray}
M_{\theta\theta'}^{\alpha}
      &=&(R^{-t}_{\theta\mu} R^{-t}_{\theta'\gamma}+\theta\leftrightarrow\theta')
    \varphi^{\text{dr}}_{\mu\gamma}
    \tfrac{\rho_\gamma f_\gamma}{\rho_{T,\mu}}
    \Delta v^{\text{eff}}_{\mu,\gamma}
    R^{t}_{\mu\alpha},
\\
M_{\theta\theta'}^{\mathfrak{f},\alpha}
      &=&(R^{-t}_{\theta\mu} R^{-t}_{\theta'\gamma}+\theta\leftrightarrow\theta')
    \varphi^{\text{dr}}_{\mu\gamma}
    \tfrac{\rho_\gamma f_\gamma}{\rho_{T,\mu}}
    \Delta a^{\text{eff}}_{\mu,\gamma}
    R^{t}_{\mu\alpha}.
\end{eqnarray}
More precisely, we made use of the following variational relations 
\begin{equation}
    \frac{\delta \rho_{\beta}}{\delta n_{\alpha}}=R^{-t}_{\beta,\alpha}\frac{n_{\alpha}}{\rho_{\alpha}}\quad,\quad \frac{\delta R^{-1}_{\alpha,\beta}}{\delta n_{\zeta}}=-T^{\rm dr}_{\alpha,\zeta}R^{-1}_{\zeta,\beta}\,.
\end{equation}
\subsection{Limit of small times and comparison with Navier-Stokes eq.  \eqref{eq:nabla_eq}}
In this section we derive the diffusive equation for density of quasiparticles emerging at initial time and we show that this is equivalent with the result of \cite{Durnin2021}.
More precisely, we consider uncorrelated local equilibrium initial states, i.e. with $E_{\theta,\theta'}(x,x';t=0)=0$.
Firsly, we solve Eq.~\eqref{eq: diffusion_full2_qp_trap_app} in normal modes $E^{(n)}$, reading 
\begin{eqnarray}
\label{eq: diffusion_full2_qp_trap_nm_app}
    &&\partial_t \big(R^{-t}E^{(n)}_{\theta\theta'}R^{-1}\big)
    + \Big[\partial_x (R^{-t}v^{\rm eff}_{\theta} E^{(n)}_{\theta\theta'}R^{-1})
    +
    \nonumber\\
    &&+ \partial_\theta (R^{-t}a^{\rm eff}_{\theta} E^{(n)}_{\theta\theta'}R^{-1})+(x,\theta\leftrightarrow x',\theta')\Big]
    =
    \nonumber\\
    &&=
    \delta(x-x')\bigl( M_{\theta\theta'}^{\;\alpha} \partial_x \rho_\alpha
    + M_{\theta\theta'}^{\mathfrak{f},\alpha} \partial_\alpha \rho_\alpha \bigr)\,,
\end{eqnarray}
where in this symbolic notation the rotation matrices are acting from left and right on the inner object.
The latter equation can be solved in the small time limit through the following ansatz
\begin{multline}
    E^{(n)}_{\theta,\theta'}(x,x',t\to 0^+)=\frac{1}{2}K^{(I)}_{\theta,\theta'}\text{sgn}(x-x')+\\+\frac{1}{2}K^{(II)}_{\theta,\theta'}\text{sgn}(x-x'+\Delta v^{\rm eff}_{\theta,\theta'}t)\,.
\end{multline}
where $K^{(I)}$ and $K^{(II)}$ are two time evolving functions. 
This solution represent the small ballistic evolution of initial fluctuation through the characteristics, that for $t\to0^+$ are not perturbed by the force (the acceleration is suppressed as $t^2$). 

In particular, the initial time condition imposes $K^{(II)}=-K^{(I)}$. For small times, this ansatz exactly solves the r.h.s of~\eqref{eq: diffusion_full2_qp_trap_nm_app}. In particular, we find 
\begin{eqnarray}
    K^{(I)}_{\theta,\theta'}(x,x) &&= \frac{1}{\Delta v^{\rm eff}_{\theta,\theta'}}\Big[\varphi^{\text{dr}}_{\theta\theta'}
    \tfrac{\rho_\theta' f_\theta'}{\rho_{T,\theta'}}
    \Delta v^{\text{eff}}_{\theta,\theta'}
    R^{t}_{\theta\alpha}\partial_x\rho_{\alpha}+\\&&\nonumber
    +\varphi^{\text{dr}}_{\theta\theta'}
    \tfrac{\rho_\theta' f_\theta'}{\rho_{T,\theta'}}
    \Delta a^{\text{eff}}_{\theta,\theta'}
    R^{t}_{\theta\alpha}\partial_\alpha\rho_{\alpha}+(\theta\leftrightarrow\theta')\Big]\,.
\end{eqnarray}
Hence, we can now write the initial time regularized correlation as
\begin{multline}
   R^{t} E^{\rm sym}_{\theta,\theta'}(x^+,x^-;t\to0^+) R= \\= \frac{1}{2}\frac{\text{sgn}(\Delta v^{\rm eff}_{\theta,\theta'})}{\Delta v^{\rm eff}_{\theta,\theta'}}\Big[\varphi^{\text{dr}}_{\theta\theta'}
    \frac{\rho_{\theta'} f_{\theta'}}{\rho_{T,\theta'}}
    \big(\Delta v^{\text{eff}}_{\theta,\theta'}
    R^{t}_{\theta\alpha}\partial_x\rho_{\alpha}+\\+
    \Delta a^{\text{eff}}_{\theta,\theta'}
    R^{t}_{\theta\alpha}\partial_\alpha\rho_{\alpha}\big)+(\theta\leftrightarrow\theta')\Big]\,.
\end{multline}
Plugging this solution into Eq.~\eqref{eq: diffusion_full1_qp_trap_app} we find four additive diffusive terms involving derivatives in space and rapidity. Firstly we consider the two terms coming from the multiplication with tensor $H$
\begin{eqnarray}
\label{eq: mult_E_by_H_app}
    &&H_\theta^{\alpha\beta}E^{\rm sym}_{\alpha,\beta}(t\to0^+)=
    \\\nonumber&&
    =R^{-t}_{\theta\gamma}\frac{\varphi^{\text{dr}}_{\gamma\mu}}{\rho_{T,\gamma}} |\Delta v^{\text{eff}}_{\mu,\gamma}|
    \big(\rho_{\mu} f_{\mu}
    \frac{\varphi^{\text{dr}}_{\gamma\mu}}{\rho_{T,\mu}}R^{t}_{\gamma\alpha}-\mu\leftrightarrow\gamma\Big)\partial_x\rho_{\alpha}
    \\\nonumber&&
    +R^{-t}_{\theta\gamma}\frac{\varphi^{\text{dr}}_{\gamma\mu}}{\rho_{T,\gamma}} \frac{\Delta a^{\text{eff}}_{\mu,\gamma}}{\text{sgn}(\Delta v^{\text{eff}}_{\mu,\gamma})}
    \big(\rho_{\mu} f_{\mu}
    \frac{\varphi^{\text{dr}}_{\gamma\mu}}{\rho_{T,\mu}}R^{t}_{\gamma\alpha}-\mu\leftrightarrow\gamma\Big)\partial_\alpha\rho_{\alpha}
    \\\nonumber&&=
    -\mathfrak{D}_{\theta}^{\gamma}\partial_x\rho_{\gamma}-\mathfrak{D}_{{\mathfrak{f}},\theta}^{\gamma}\partial_\gamma\rho_{\gamma}\,,
\end{eqnarray}
where we used the definition of inverse rotation matrix $R^{-t}R^{t}=1$ and where the factor $1/2$ is compensated by the symmetry of $H$.
Similarly we now consider the two terms coming from the multiplication of $E^{\rm sym}(t=0^+)$ with the tensor $H^{\mathfrak{f}}$
\begin{eqnarray}
\label{eq: mult_E_by_Hf_app}
    &&H_\theta^{\mathfrak{f},\alpha\beta}E^{\rm sym}_{\alpha,\beta}(t\to0^+)=
    \\\nonumber&&=R^{-t}_{\theta\gamma}\frac{\varphi^{\text{dr}}_{\gamma\mu}}{\rho_{T,\gamma}} \frac{\Delta a^{\text{eff}}_{\mu,\gamma}}{\text{sgn}(\Delta v^{\text{eff}}_{\mu,\gamma})}
    \big(\rho_{\mu} f_{\mu}
    \frac{\varphi^{\text{dr}}_{\gamma\mu}}{\rho_{T,\mu}}R^{t}_{\gamma\alpha}-\mu\leftrightarrow\gamma\Big)\partial_x\rho_{\alpha}
    \\\nonumber&&
    +R^{-t}_{\theta\gamma}\frac{\varphi^{\text{dr}}_{\gamma\mu}}{\rho_{T,\gamma}}\frac{(\Delta a^{\text{eff}}_{\mu,\gamma})^2}{|\Delta v^{\text{eff}}_{\mu,\gamma}|}
    \big(\rho_{\mu} f_{\mu}
    \frac{\varphi^{\text{dr}}_{\gamma\mu}}{\rho_{T,\mu}}R^{t}_{\gamma\alpha}-\mu\leftrightarrow\gamma\Big)\partial_\alpha\rho_{\alpha}    
    \\\nonumber&&=
    -\mathfrak{D}_{\mathfrak{f},\theta}^{\gamma}\partial_x\rho_{\gamma}-\mathfrak{D}_{\mathfrak{f}^2,\theta}^{\gamma}\partial_\gamma\rho_{\gamma}\,.
\end{eqnarray}
Finally, using the results~\eqref{eq: mult_E_by_H_app} and \eqref{eq: mult_E_by_Hf_app} into~\eqref{eq: diffusion_full1_qp_trap_app} we find the diffusion equation emerging at $t\to0^+$ for an integrable systems in inhomogeneous fields quenched from an uncorrelated local equilibrium state
\begin{multline}
\label{eq: diffusion_full1_qp_trap_app_final}
    \partial_t \rho_\theta
    + \partial_x (v^{\text{eff}}_\theta \rho_\theta)
    + \partial_\theta (a^{\text{eff}}_\theta \rho_\theta)=\frac{1}{2}\partial_x\mathfrak{D}_{\theta}^{\gamma}\partial_x\rho_{\gamma}+\\+\frac{1}{2}\partial_x\mathfrak{D}_{{\mathfrak{f}},\theta}^{\gamma}\partial_\gamma\rho_{\gamma}+\frac{1}{2}\partial_{\theta}\mathfrak{D}_{\mathfrak{f},\theta}^{\gamma}\partial_x\rho_{\gamma}+\frac{1}{2}\partial_{\theta}\mathfrak{D}_{\mathfrak{f}^2,\theta}^{\gamma}\partial_\gamma\rho_{\gamma}\,,
\end{multline}
being in perfect agreement with the result presented in \cite{Durnin2021}.
\subsection{Derivation of Eq.~\eqref{eq: H_matr_app} and~\eqref{eq: Hf_matr_app}}
\label{sec: derivation_H_matrix_app}
In this section we show the steps to derive the tensors shown in Eq.~\eqref{eq: H_matr_app} and~\eqref{eq: Hf_matr_app}. Since the derivation in the two cases is completely analogous, we will only show how to compute $H$, meanwhile the steps for $H^{\mathfrak{f}}$ will straightforwardly follow the same logic.
Hence, we aim to compute the following variation
\begin{equation}
    H_{\theta}^{\theta'\theta''}=\frac{\delta A_{\theta}^{\theta'}}{\delta \rho_{\theta''}}=\frac{\delta A_{\theta}^{\theta'}}{\delta n_{\gamma}}\frac{\delta n_{\gamma}}{\delta \rho_{\theta''}}\,.
\end{equation}
Thus, we consider
\begin{multline}
    \frac{\delta A_{\theta}^{\theta'}}{\delta n_{\theta''}} = \Big[\frac{\delta R^{-t}_{\theta,\gamma}}{\delta n_{\theta''}}v^{\rm eff}_{\gamma} R^t_{\gamma,\theta'} + R^{-t}_{\theta,\gamma}v^{\rm eff}_{\gamma} \frac{\delta R^t_{\gamma,\theta'}}{\delta n_{\theta''}}\Big]+\\+\Big[R^{-t}_{\theta,\gamma} \frac{\delta v^{\rm eff}_{\gamma}}{\delta n_{\theta''}}R^t_{\gamma,\theta'}\Big] \equiv W_{\theta}^{\theta'\theta''}+Z_{\theta}^{\theta'\theta''}\,,
\end{multline}
where the tensors $W$ and $Z$ are respectively defined by the arguments inside the first and second square bracket.
We now compute separately the two contributions
using the useful relations \cite{doyon_lecture_2019, Medenjak2020, Biagetti2025}
\begin{gather}
    \frac{\delta n_{\theta}}{\delta \rho_{\theta'}} = \frac{R^t_{\theta,\theta'}} {\rho_{T,\theta}}\quad,\quad \frac{\delta R^{-1}_{\theta,\theta'}}{\delta n_{\gamma}}=-\varphi^{\rm dr}_{\theta,\gamma}R^{-1}_{\gamma,\theta'}\,,
    \\
    \frac{\delta v^{\rm eff}_{\theta}}{\delta n_{\theta'}} = - \frac{\rho_{T,\theta'}}{ \rho_{\theta}}\Delta v_{\theta,\theta'}R^{-t}_{\theta,\theta'}\,.
\end{gather}
They explicitly read
\begin{eqnarray}
\label{eq: Z_term_app}
    Z_{\theta}^{\theta'\theta''} &=&R^{-t}_{\theta\gamma} \frac{\rho_{T,\theta''}}{ \rho_{\gamma}}\Delta v_{\theta''\gamma}R^{-t}_{\gamma\theta''}R^t_{\gamma\theta'}\,,
\\
\label{eq: W_term_app}
    W_{\theta}^{\theta'\theta''} &=&R^{-t}_{\theta\theta''}\varphi^{\rm dr}_{\theta''\gamma}
    \Delta v^{\rm eff}_{\gamma\theta''} R^t_{\gamma\theta'}\,
\\
    H_{\theta}^{\theta'\theta''} &=&\Big(W_{\theta}^{\theta'\gamma}+Z_{\theta}^{\theta'\gamma}\Big)\frac{R^t_{\gamma,\theta''}}{\rho_{T,\theta''}} = 
    \\\nonumber&=&R^{-t}_{\theta\gamma}
    \Delta v^{\text{eff}}_{\mu,\gamma}\tfrac{\varphi^{\text{dr}}_{\gamma\mu}}{\rho_{T,\gamma}}
    (R^{t}_{\mu\theta'} R^{t}_{\gamma\theta''}+R^{t}_{\mu\theta''} R^{t}_{\gamma\theta'})\,,
\end{eqnarray}
being the result already shown in~\eqref{eq: H_matr_app}.
\subsection{Derivation of Eq.~\eqref{eq: der_Bmatr_app} and~\eqref{eq: der_Bmatr_f_app}}
\label{eq: derivation_M_matrix_app}
In this section we derive Eq.~\eqref{eq: der_Bmatr_app} and~\eqref{eq: der_Bmatr_f_app}. In particular, Eq.~\eqref{eq: der_Bmatr_app} is a trivial consequence of the fact that the only independent space varying functions in the theory are $\rho(x)$ and $V(x)$, which enters in the effective velocity through the derivative of the energy density. Hence, the space derivative of any object can be simply written as variations with respect to these two functions.
On the other hand, the argument to show~\eqref{eq: der_Bmatr_f_app} is more involved, since in principle the rapidity is also entering non-trivially in the rotation matrix through the scattering shift $\varphi$. In this section we specialize to Galilean invariant systems where the scattering shift depends only on the difference of rapidities $\varphi_{\theta\theta'}=\varphi(\theta-\theta')$ and $\rho_{T,\theta}=1^{\rm dr}_{\theta}$.
We now derive the some useful relation for the rapidity derivatives. Firstly we compute the derivative of 
the filling function $n_{\theta}$
\begin{eqnarray}
    \nonumber
    &&\partial_{\theta} n_{\theta} 
    =\frac{\partial_{\theta}\rho_{\theta}}{1^{\rm dr}_{\theta}}-\frac{n_{\theta}}{1^{\rm dr}_{\theta}}\varphi^{\rm dr}_{\theta,\gamma}1^{\rm dr}_{\gamma}\partial_{\gamma}n_{\gamma}\,,
    \\\nonumber
    &&(\delta_{\theta\gamma}+n_{\theta}\varphi^{\rm dr}_{\theta\gamma})1^{\rm dr}_{\gamma}\partial_{\gamma}n_{\gamma}=\partial_{\theta}\rho_{\theta}\,,
    \\
    &&\partial_{\theta}n_{\theta}=\frac{R^{t}_{\theta\gamma}}{1^{\rm dr}_{\theta}}\partial_{\gamma}\rho_{\gamma}\,,
\end{eqnarray}
where in the first line we used $R^{-1}_{\theta\gamma}=\delta_{\theta,\gamma}+\varphi^{\rm dr}_{\theta\gamma}n_{\gamma}$ and that $1^{\rm dr}_{\theta}=R^{-1}_{\theta\gamma} 1_{\gamma}$.
We now can compute the derivative of a generic dressed function $f^{\rm dr}_{\theta}$, where $f_{\theta}$ is a  state independent function of rapidity
\begin{equation}
    \partial_{\theta}f^{\rm dr}_{\theta}=R^{-1}_{\theta\gamma}\partial_{\gamma}f_{\gamma}+\varphi^{\rm dr}_{\theta\gamma}\frac{f^{\rm dr}_{\gamma}}{1^{\rm dr}_{\gamma}}R^{t}_{\gamma\alpha}\partial_{\alpha}\rho_{\alpha}\,,
\end{equation}
in which we mainly used the property $\partial_{\theta}\varphi_{\theta\theta'}=-\partial_{\theta'}\varphi_{\theta\theta'}$ and integration by parts. In particular, this  crucially permits to compute the derivatives of the dressing function
\begin{eqnarray}
    \partial_{\theta}R^{-t}_{\theta\theta'}&=&R^{-t}_{\theta\theta'}\partial_{\theta'}+R^{-t}_{\theta\gamma}\frac{T^{\rm dr}_{\gamma\theta'}}{1^{\rm dr}_{\gamma}}R^{t}_{\gamma\alpha}\partial_{\alpha}\rho_{\alpha}\,,
    \\
    \partial_{\theta}R^{-1}_{\theta\theta'}&=&-\partial_{\theta'}R^{-1}_{\theta,\theta'}+T^{\rm dr}_{\theta,\gamma}\frac{R^{t}_{\gamma,\alpha}}{1^{\rm dr}_{\gamma}}\partial_{\alpha}\rho_{\alpha}R^{-1}_{\gamma,\theta'}\,.
\end{eqnarray}
We finally can compute the derivative of the matrix $B^{\mathfrak{f}}$ defined in Eq.~\eqref{eq: def_B_f_app}
\begin{eqnarray}
    &&(\partial_{\theta}+\partial_{\theta'})B^{\mathfrak{f}}
    =R^{-t}_{\theta,\gamma}f_{\gamma}\partial_{\gamma}(a^{\rm eff}_{\gamma}\rho_{\gamma})R^{-t}_{\gamma,\theta'}+
    \\\nonumber&&+(R^{-t}_{\theta\beta}R^{-t}_{\theta'\gamma}+R^{-t}_{\theta\gamma}R^{-t}_{\theta'\beta})a^{\rm eff}_{\gamma}\rho_{\gamma}f_{\gamma}\frac{\varphi^{\rm dr}_{\gamma\beta}}{1^{\rm dr}_{\beta}}R^{t}_{\beta\alpha}\partial_{\alpha}\rho_{\alpha}
    \\\nonumber&&+R^{-t}_{\theta,\gamma}a^{\rm eff}_{\gamma}\rho_{\gamma}f_{\gamma}\partial_{\theta'}(R^{-1}_{\gamma,\theta'}-R^{-1}_{\gamma,\theta'})\,,
\end{eqnarray}
where the last line vanishes. Hence, we observe that the last equation is equivalent to~\eqref{eq: der_Bmatr_f_app}.

\bibliography{apssamp.bib}
\end{document}